\documentclass[aps, prd, reprint, superscriptaddress, nofootinbib]{revtex4-2}

\usepackage{amssymb,amsmath,verbatim,mathtools,needspace,enumitem,etoolbox,graphicx,microtype,afterpage,bm,subcaption,siunitx,booktabs,multirow}

\usepackage[unicode, colorlinks=true, linkcolor=linkcolor, citecolor=linkcolor, filecolor=linkcolor,urlcolor=linkcolor, pdfusetitle]{hyperref}

\usepackage[T1]{fontenc}
\usepackage[utf8]{inputenc}
\usepackage{tabularx}
\newcolumntype{Y}{>{\raggedright\arraybackslash}X}
\usepackage[dvipsnames, usenames]{xcolor}
\definecolor{linkcolor}{rgb}{0.1,0.2,0.6}
\usepackage[italicdiff]{physics}
\usepackage{float}
\usepackage{subcaption}
\usepackage{etoolbox}

\AtBeginEnvironment{figure}{%
    \captionsetup{justification=justified, singlelinecheck=false}%
}
\AtBeginEnvironment{table}{%
    \captionsetup{justification=justified, singlelinecheck=false}%
}

\begin{document}
\title{Tracing the Trace Anomaly of Dense Matter inside Neutron Stars}

\author{Shiyue Ren}
\email[]{rsy@mail.nankai.edu.cn}
\affiliation{School of Physics, Nankai University, Tianjin 300071, China} 
\author{Lap-Ming Lin}
\email[]{lmlin@cuhk.edu.hk}
\affiliation{Department of Physics, The Chinese University of Hong Kong, Hong Kong SAR, China}

	\date{\today}
	
	\begin{abstract}
		\begin{center}
			ABSTRACT
		\end{center}
The trace anomaly $\Delta$ is an important quantity that measures the broken conformal symmetry in neutron star matter. In this work, we present quasi-universal relations that connect the stellar profile of $\Delta$ to the compactness, moment of inertia, and tidal deformability of neutron stars.
We apply the quasi-universal relations to determine the trace anomaly profiles for PSR J0030+0451
and PSR J0740+6620 based on their mass-radius measurements. We also analyze PSR J0737-3039A according to its moment of inertia inferred from Bayesian modeling of nuclear equation of state. A recent multimessenger constraint on the tidal deformability is also studied, resulting in an estimate value of the trace anomaly $\Delta_c = 0.1770^{+0.0365}_{-0.0432}$ at the center of a $1.4M_\odot$ canonical neutron star.
It is expected that more precise observations from both electromagnetic and gravitational-wave channels in the future will provide tighter constraints on the behavior of $\Delta$ inside neutron stars.
	\end{abstract}
	
	\maketitle

\section{Introduction}\label{sec:introduction}

Neutron stars (NSs) are remnants of massive stars that have experienced core collapse supernovae after exhausting their nuclear fuel~\cite{Shapiro2004}. 
These incredibly dense stars are among the most fascinating objects in the universe. 
With observed masses ranging from about 1 to 2 times that of the Sun but radii of only about 10 km, the
cores of NSs exhibit an extreme high density environment reaching a few times or beyond the nuclear
saturation density. For this reason, since their discovery more than half a century ago, NSs have long been regarded as unique laboratories for testing and constraining theories of nuclear-matter equation of state (EOS)~\cite{LATTIMER2007109}.

Although the nuclear-matter EOS relevant to the high density core of NSs is still largely uncertain, 
interesting NS observations in the past decade, such as the discovery of NSs with masses 
$M \approx 2 M_\odot$ \cite{demorestTwosolarmassNeutronStar2010,Antoniadis_2013} and more recently the 
mass-radius observations made by Neutron Star Interior Composition Explorer (NICER) \cite{Miller_nicer2019,Riley_2019,Miller_nicer2021,Riley_2021}, have provided important information to constrain theoretical models. 
An important quantity to characterize an EOS is the speed of sound (squared) $c_s^2 = dP/d\rho$, where $P$ and $\rho$ are the pressure and energy density, respectively. 
The conditions of thermodynamical stability and causality restrict the speed of sound to the range
$0 \leq c_s^2 \leq 1$, where the speed of light $c \equiv 1$.  
At asymptotic high densities, it is expected that $c_s^2$ approaches the so-called conformal limit $c_s^2=1/3$ 
where quantum chromodynamics (QCD) regains conformal symmetry. However, in the non-perturbative regime 
relevant to the density range of neutron stars, $c_s^2$ may exceed the conformal limit and the upper bound 
is set only by the causality limit $c_s^2=1$. In fact, the existence of NSs with masses $M\approx 2 M_\odot$
poses a significant challenge to the conjecture that $c_s^2 \leq 1/3$ at all densities
\cite{bedaqueSoundVelocityBound2015}. 
The variation of $c_s^2$ with densities and how it approaches the conformal limit have gained a lot of attention (e.g., \cite{Tews_2018,McLerran_2019,Altiparmak_2022,Tan_2022,Annala_2023}). 
It has also been proposed that $c_s^2$ can approach the conformal limit at about $2 - 3$ times 
the nuclear saturation density, but the conformal symmetry is not restored in the so-called pseudoconformal model~\cite{Paeng_2017,MaRho_2019}.

It has recently been suggested that the normalized QCD trace anomaly $\Delta$ is a more comprehensive quantity than $c_s^2$ and provides a new measure of conformality \cite{fujimotoTraceAnomalySignature2022}
(see, e.g., \cite{Marczenko_2023,Jimenez_2024,Cai_2025} for some recent work). 
In massless QCD, conformal symmetry implies that the trace of the energy-momentum tensor vanishes. However,
due to the running of the coupling constant and the presence of mass terms, this symmetry is broken, leading to a non-zero trace. In \cite{fujimotoTraceAnomalySignature2022}, it is proposed to measure the trace anomaly by defining 
\begin{equation}
\Delta \equiv { {\rho - 3 P} \over {3\rho} }  , 
\end{equation}
where the numerator in the definition is the trace of the energy-momentum tensor and the denominator is a normalization factor used to make $\Delta$ dimensionless\footnote{We shall simply refer to $\Delta$ as the trace anomaly rather than the more appropriate term, the normalized trace anomaly.}.
It is noted that $\Delta=0$ at asymptotically high densities, where QCD regains conformal symmetry. 
However, in the density range relevant to NSs, $\Delta$ is only subject to the 
conditions of thermodynamic stability ($P>0$) and causality ($P \leq \rho$), leading to the allowed range 
$-2/3 \leq \Delta < 1/3$. The speed of sound can be expressed as 
$c_s^2 =1/3 - \Delta - \rho d\Delta/d\rho$~\cite{fujimotoTraceAnomalySignature2022}, and hence $\Delta = 0$ and $c_s^2=1/3$ in the conformal limit.
In addition to proposing $\Delta$ as a new measure of conformality, the authors of \cite{fujimotoTraceAnomalySignature2022} also conjecture that $\Delta \geq 0$ in the interiors of NSs. However, recent studies suggest that $\Delta$ may cross zero and become negative
\cite{Brandes_2025,Fukushima_2025}.
In this work, we study the correlations between $\Delta$ and various global quantities of NSs, with the aim
of exploring the properties of $\Delta$, and hence the dense matter of NSs from observation data. 
Specifically, we shall consider the compactness, moment of inertia, and tidal deformability of NSs
because of their great observational relevance and theoretical importance. 

The compactness of a NS is an important quantity because of its connection to the mass and radius of 
the star. It can be obtained directly by measuring the gravitational redshift of the spectral lines at the 
surface of a NS \cite{cottamGravitationallyRedshiftedAbsorption2002}. It can also be obtained if the mass and radius of a NS can be measured simultaneously, as demonstrated by recent mass-radius observations for PSR J0740+6620  \cite{Miller_nicer2021,Riley_2021} and 
PSR J0030+0451 \cite{Miller_nicer2019,Riley_2019} by NICER mission. 
On the other hand, the second quantity we considered, the moment of inertia, actually contains richer information about the mass distribution inside a NS, potentially providing more information about the EOS. 
It can be measured through the spin-orbit coupling on the orbital motion of binary pulsar systems 
\cite{kramerDoublePulsarSystem2009}. Specifically, the moment of inertia of PSR J0737-3039A is expected to be measurable to about 10\% accuracy \cite{lattimerConstrainingEquationState2005,Hu_Huanchen2020}. 

The tidal deformability of NSs has gained a lot of attention in the era of gravitational wave astronomy. 
It quantifies the deformation of a NS under the influence of an external tidal field, such as that 
exerted by a companion star in the late inspiral phase of a binary NS merger \cite{flanaganConstrainingNeutronstarTidal2008}. 
The first gravitational wave signal from a binary NS system GW170817~\cite{abbottGW170817ObservationGravitational2017} has been used to constrain
the tidal deformability (see \cite{Chatziioannou2020} for a review). 
Future observations of binary NS events are expected to offer tighter constraints on the tidal deformability, and provide more insights on the dense matter EOS. 

In addition to their observational significance, the compactness, moment of inertia, and tidal deformability 
are also connected through the I-Love-Q \cite{yagiILoveQRelationsNeutron2013} and I-C universal relations \cite{Jiang_2020}. These relations are approximately insensitive to the EOS, exhibiting about a 1\% to 10\% insensitivity for the former and latter relations, respectively. 
In this paper, we establish quasi-universal relations that connect the profile of trace anomaly to these 
important NS observables. Specifically, our main result (Eq.~(\ref{eq:fit})) can be used to determine the ratio $X$ between pressure and energy density, which is directly related to the trace anomaly 
(see Eq.~(\ref{eq:Xtrace})), to within about 10\% level when any one of these dimensionless NS quantities is given.

The plan of the paper is as follows. In Sec.~\ref{sec:method}, we briefly outline the methods for 
computing the relevant NS quantities in this study, and present the mass-radius relations of our chosen EOS models. In Sec.~\ref{sec:results}, we present the quasi-universal relations for the profile of $X$ and some 
validation tests. Sec.~\ref{sec:apply} discusses the applications of the quasi-universal relations. Finally, 
we summarize and discuss the implications of our study in Sec.~\ref{sec:discuss}. Unless otherwise noted, we use units where $c=G=1$.

\section{Method and EOS models}
\label{sec:method}

In this work, we study the relationships between the trace anomaly $\Delta$ and the compactness $C=M/R$, the moment of inertia $I$, and the tidal deformability $\lambda$ of NSs, where $M$ and $R$ are the mass and radius of the stars, respectively. For the latter two quantities, we consider the dimensionless quantities ${\bar I} \equiv I/M^3$ and $\Lambda \equiv \lambda / M^5$, which were used in the I-Love-Q universal relations~\cite{yagiILoveQRelationsNeutron2013}. 
As these physical quantities have been well studied, we only outline the formulation
for their computations and refer the reader to the literature for more details. 
For a given EOS model that relates the pressure $P$ and energy density $\rho$ of a perfect fluid, 
the spacetime metric functions and matter profiles, such as $P(r)$ and $\rho(r)$, of a nonrotating NS in general relativity are determined by solving the Tolman-Oppenheimer-Volkoff (TOV) equation \cite{tolmanStaticSolutionsEinsteins1939}. The mass $M$ and radius $R$ of the star are determined by providing a central energy density $\rho_c$ in the computation. The resulting stellar configuration then serves as an unperturbed background solution for the computation of $I$ and $\lambda$. 


\begin{figure}[H]
    \includegraphics[width=\columnwidth]{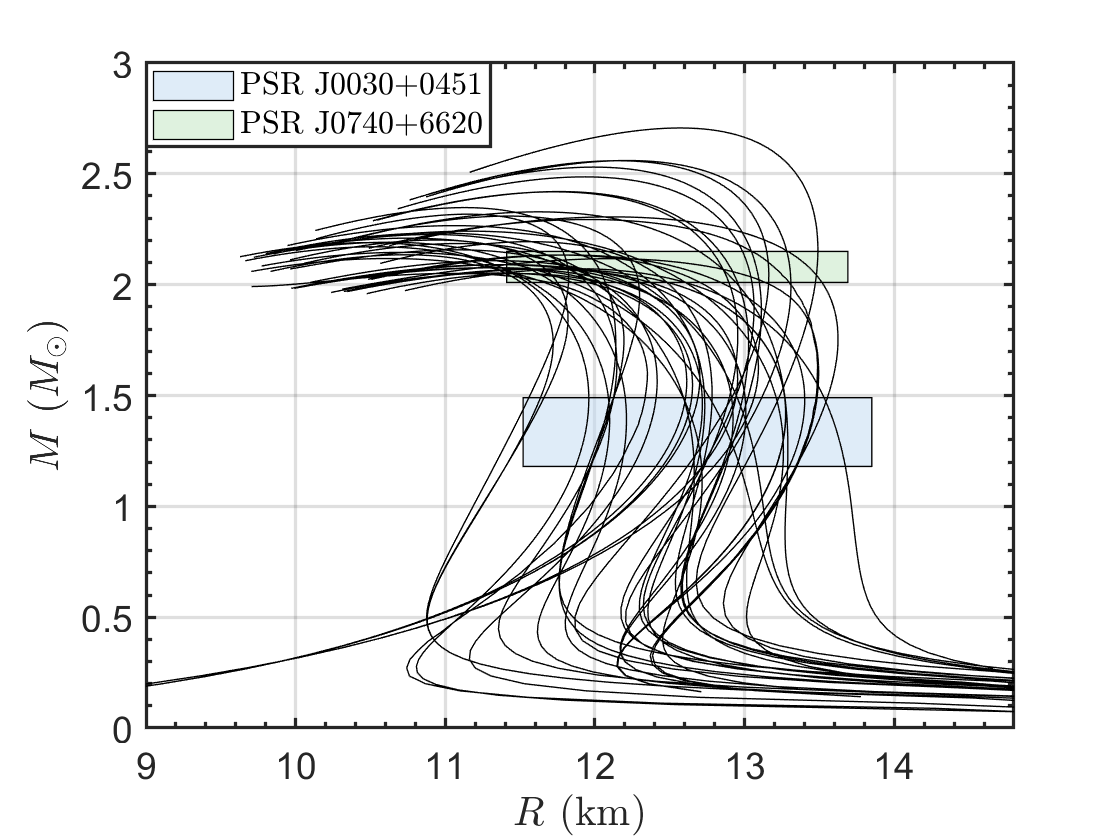}
    \caption{Mass-radius relations of the EOS models employed in this work (see Appendix~\ref{sec:EOS_models}). 
    The rectangular boxes represent the observational estimates for PSR J0740+6620~\cite{Riley_2021} and PSR J0030+0451~\cite{Riley_2019}. }
\label{fig:MRtest}
\end{figure}

The moment of inertia of a slowly rotating NS is calculated perturbatively by expanding $I=J/\Omega$ to first order in $\Omega$, where $J$ and $\Omega$ are the angular momentum and angular velocity of the star, respectively \cite{hartleSlowlyRotatingRelativistic1967}. On the other hand, the tidal deformability $\lambda$ of a nonrotating NS in the
static tide limit is defined by $Q_{ij} \equiv - \lambda {\cal E}_{ij}$, where $Q_{ij}$ is the traceless quadrupole moment tensor of the star that characterizes its deformation due to the tidal field tensor ${\cal E}_{ij}$ created by a companion star in a binary system \cite{hindererTidalLoveNumbers2008}. 

To study the level of EOS sensitivity for the relations among $\Delta$, $C$, ${\bar I}$, and $\Lambda$, we chose 45 cold NS EOS models from the CompOSE database~\cite{typelCompOSECompStarOnline2013} which are listed 
in Table~\ref{tab:eos_list_row_sorted} in Appendix~\ref{sec:EOS_models}. 
These models are chosen to be consistent with the mass-radius observations for PSR J0740+6620~\cite{Riley_2021} and PSR J0030+0451 \cite{Riley_2019} made recently by NICER. 
In Fig.~\ref{fig:MRtest}, we plot the mass-radius relations of the chosen EOS models together with the estimates ($12.39^{+1.30}_{-0.98}~\rm{km}$, $2.072^{+0.067}_{-0.066}~M_\odot$) for PSR J0740+6620 \cite{Riley_2021} and ($12.71_{-1.19}^{+1.14}~\rm{km}$, $1.34_{-0.16}^{+0.15}~M_\odot$) for PSR J0030+0451 \cite{Riley_2019}. The EOS models are chosen randomly in such a way that their mass-radius relations
can cover the observational bounds, represented by the rectangular boxes, of the two NSs uniformly as shown in the figure.

\section{Results}
\label{sec:results}
 \begin{widetext}
\begin{figure}[H]
\centering
  \begin{minipage}{0.33\linewidth}
  \includegraphics*[width=6.3cm]{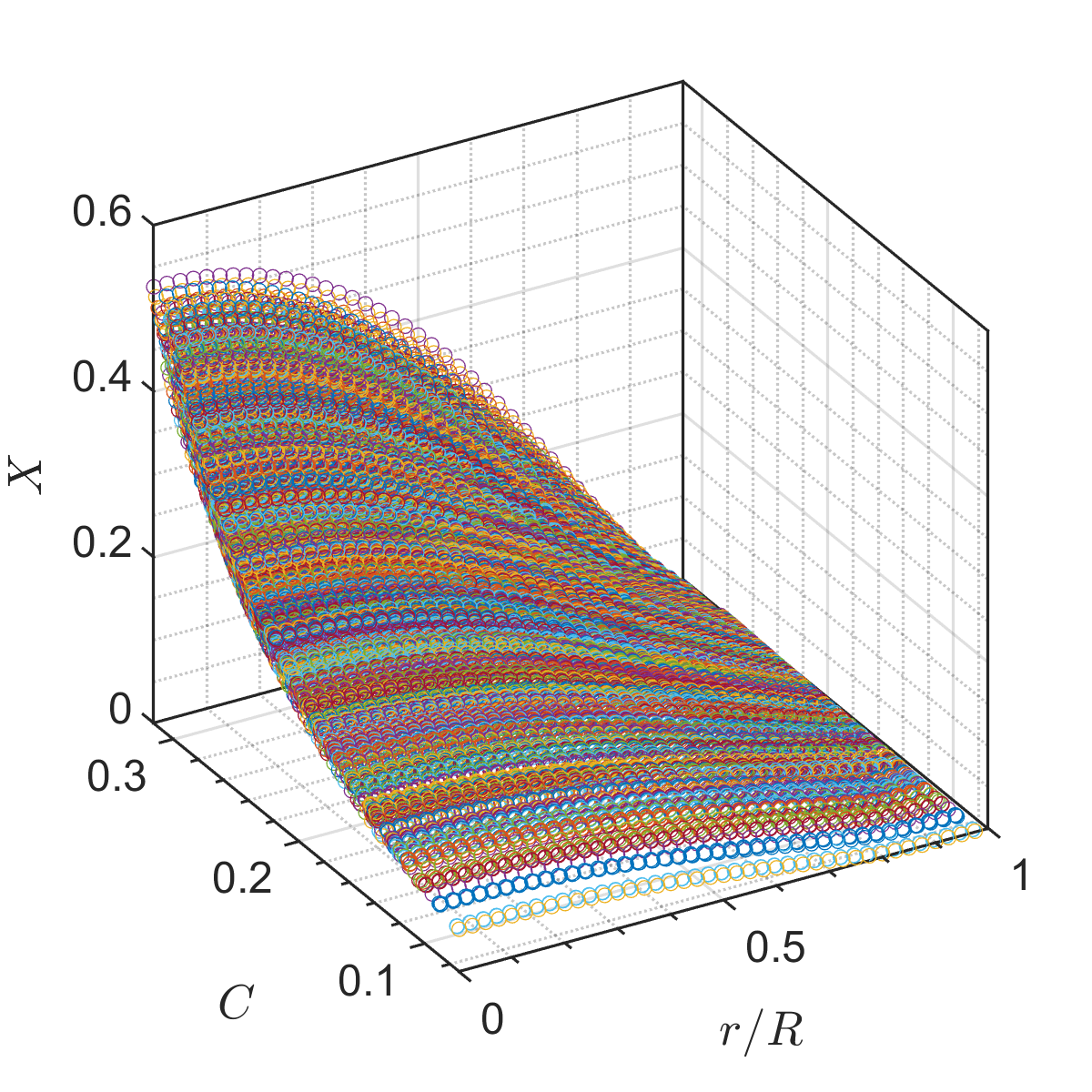}
  \includegraphics*[width=6.3cm]{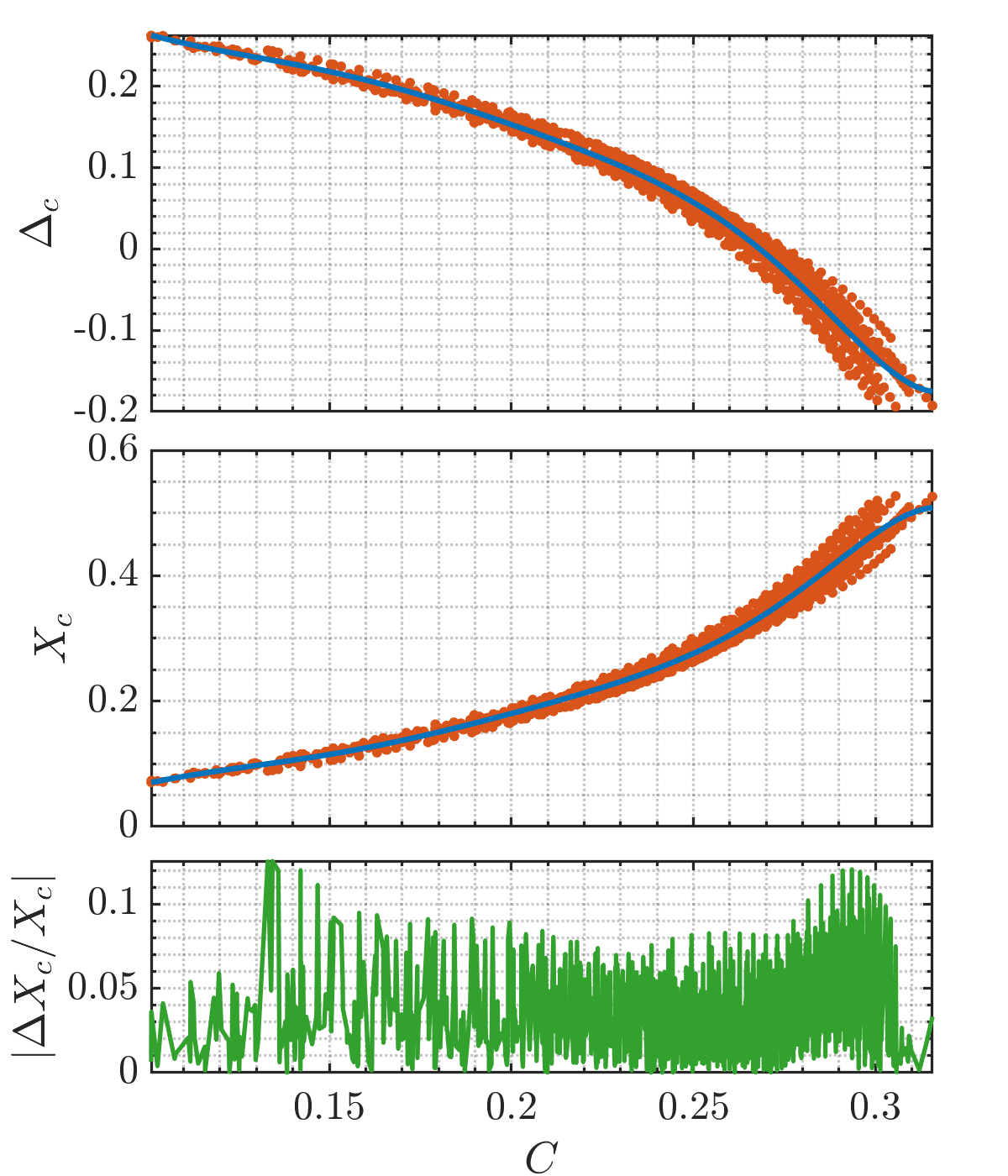}
  \end{minipage}%
  \begin{minipage}{0.33\linewidth}
  \includegraphics*[width=6.3cm]{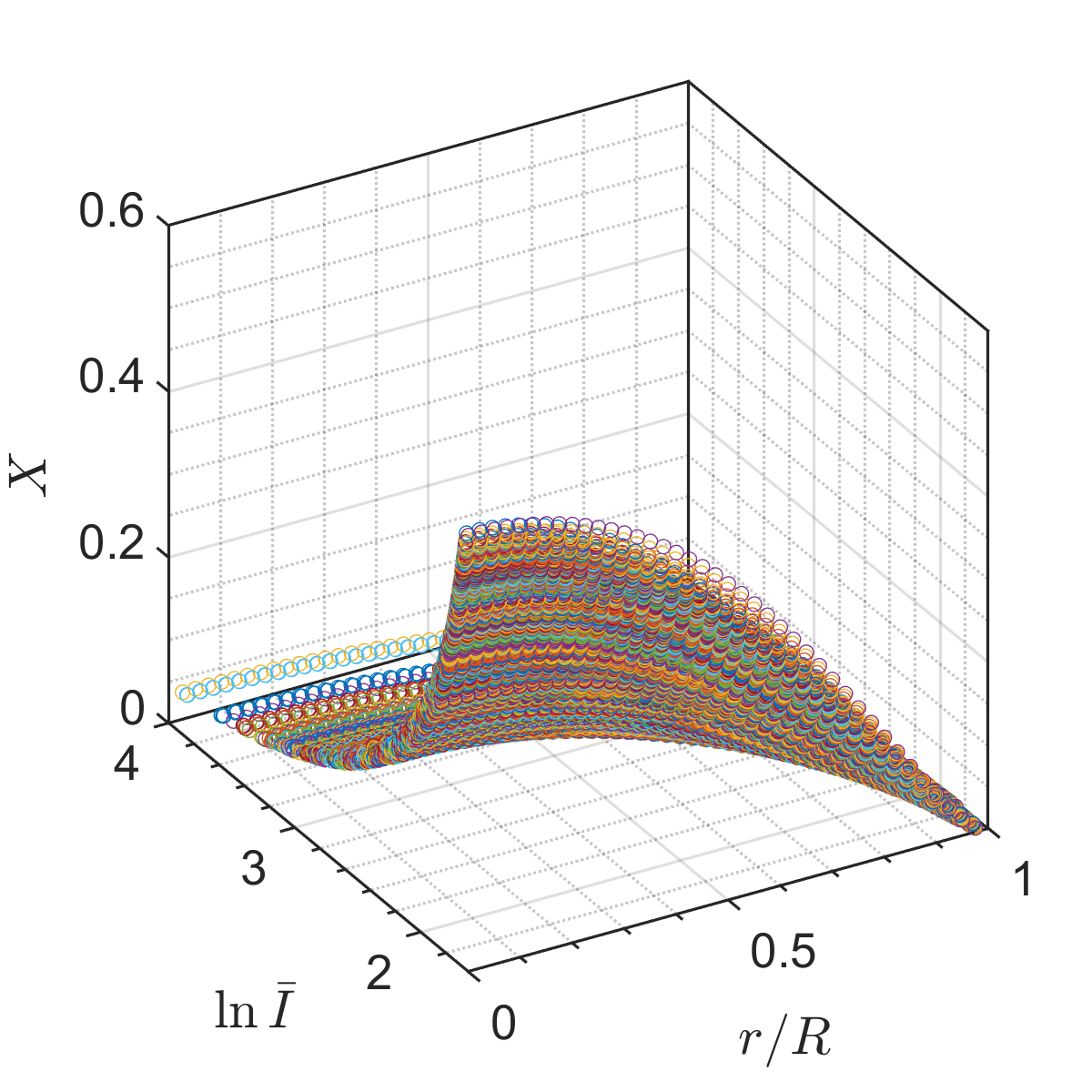}
  \includegraphics*[width=6.3cm]{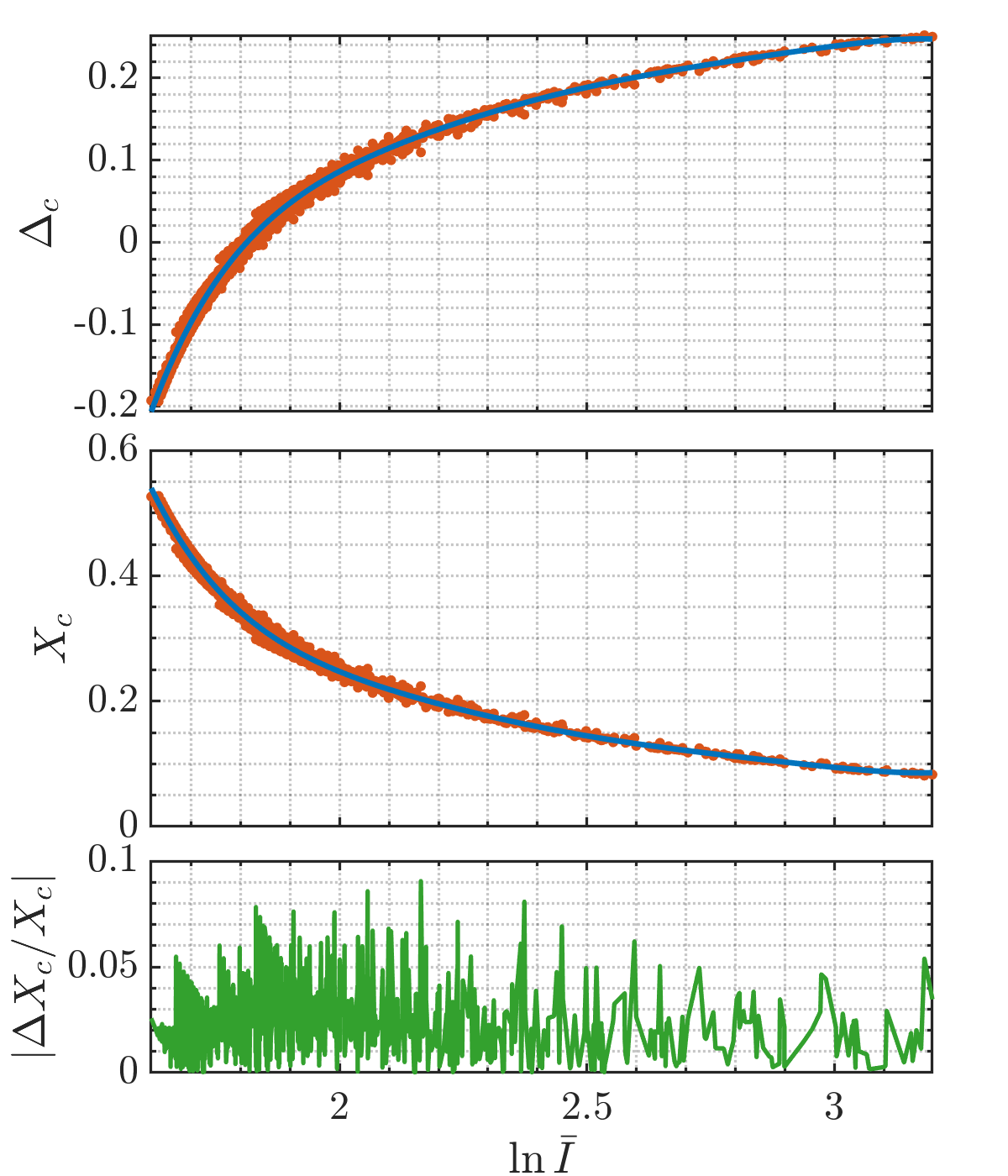}
 \end{minipage}%
  \begin{minipage}{0.33\linewidth}
  \includegraphics*[width=6.3cm]{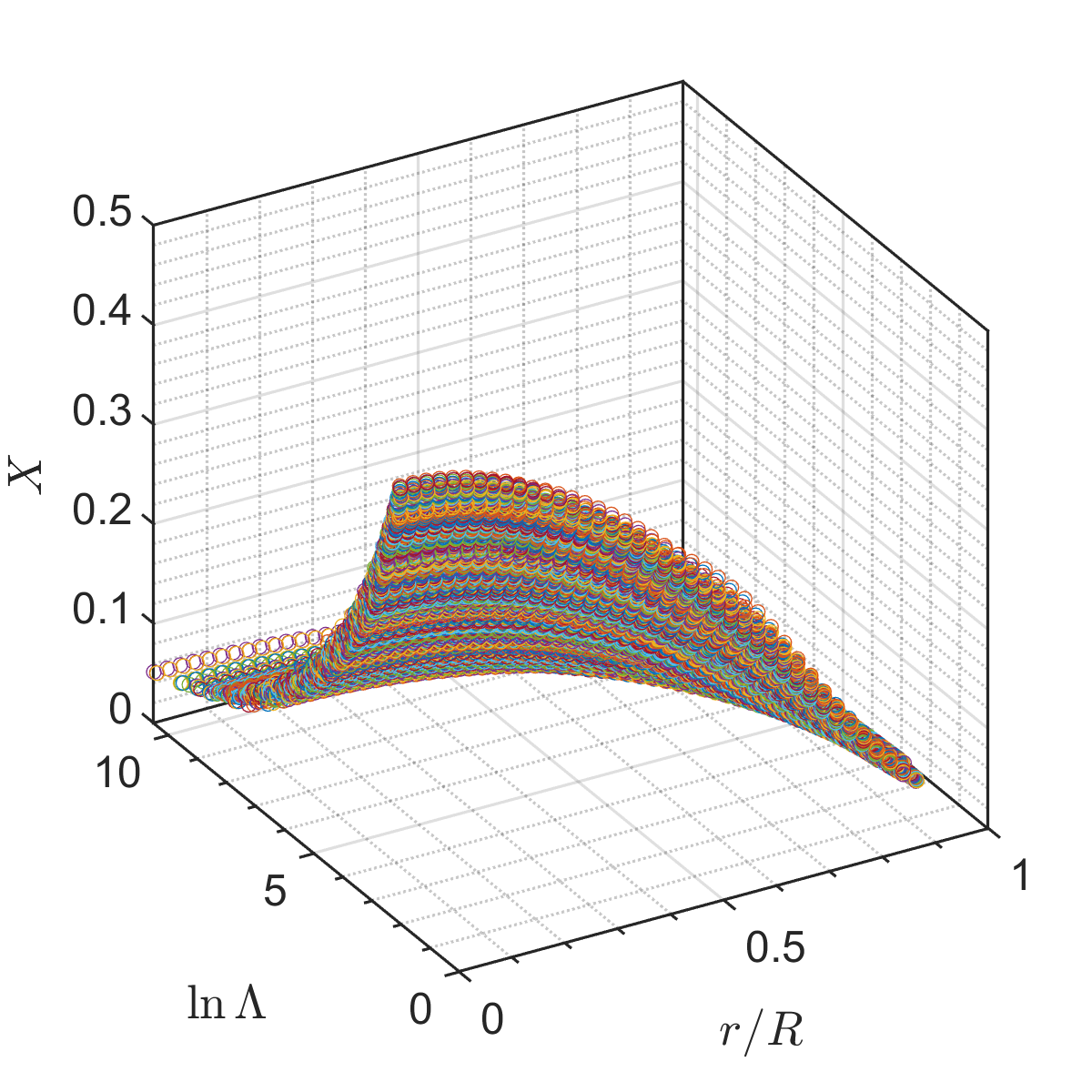}   
  \includegraphics*[width=6.3cm]{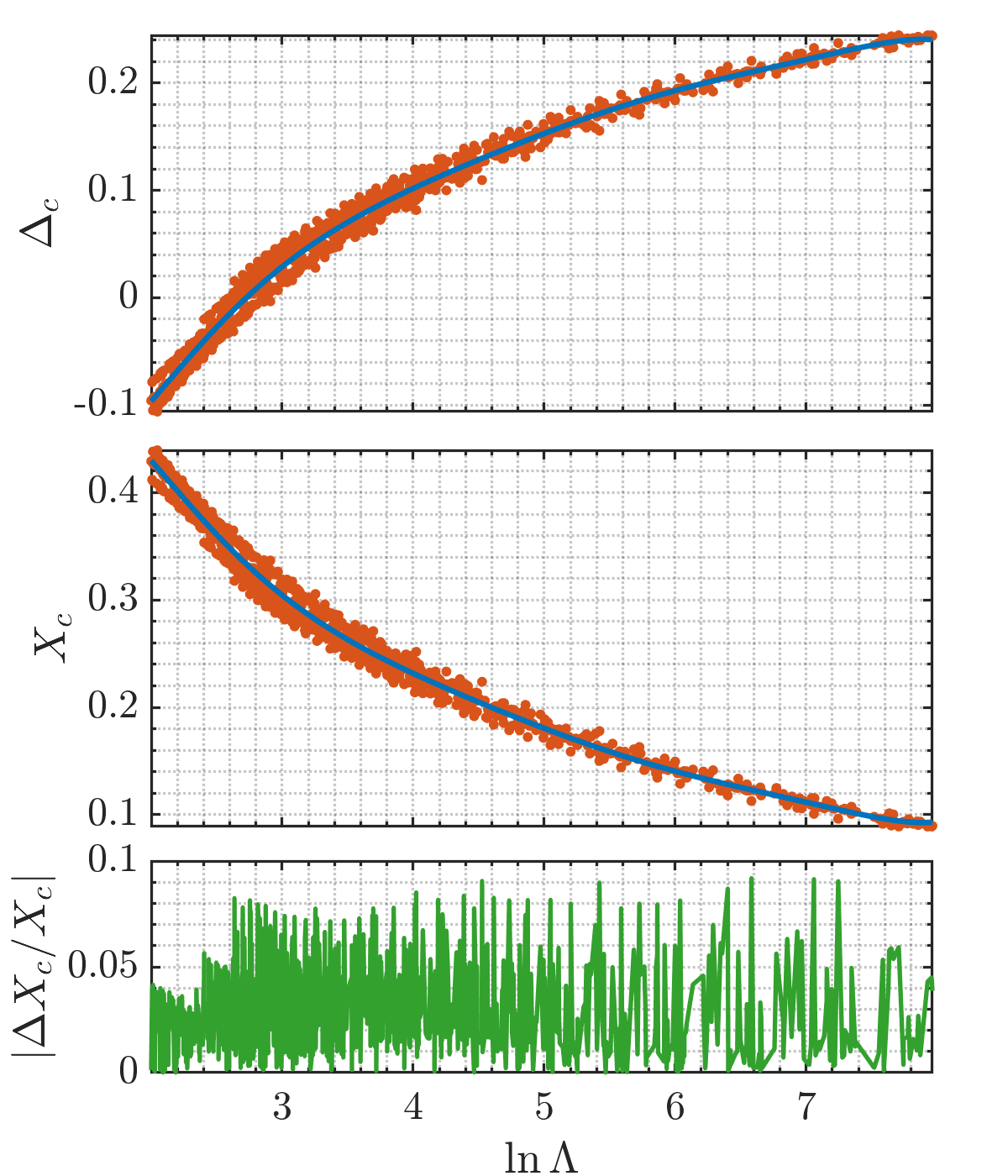}
  \end{minipage}%
\captionsetup{justification=justified, singlelinecheck=false}
 \caption{ First row: $X=P/\rho$ as a function of $r/R$ and $C$ (left panel), $\ln {\bar I}$ (middle panel), and $\ln {\Lambda}$ (right panel). Second row: The central value of the trace anomaly $\Delta_c$ is plotted against $C$, $\ln {\bar I}$, and $\ln \Lambda$. Third row: Similar to the second row, but for the central 
 value of $X$. The blue line in each panel is obtained from Eq.~(\ref{eq:fit}) evaluated at the center $r=0$. 
 Fourth row: The relative differences between the numerical data $X_c$ and the blue lines in the third row.  }
 \label{fig:3d_plots}
\end{figure}
\end{widetext}


\subsection{Quasi-universal relations}


In this section, we study the correlations between the profile of the trace anomaly $\Delta$ inside a NS and 
its global quantities $C$, $\bar I$, and $\Lambda$. 
To parameterize the radial positions in a dimensionless manner, it is natural to introduce the normalized coordinate $z\equiv r/R$, where $R$ is the stellar radius. For convenience, we use the function 
\begin{equation}
    X(z) \equiv {P \over \rho} ={1 \over 3}-\Delta(z) ,
\label{eq:Xtrace}
\end{equation}
as a proxy for the trace anomaly function $\Delta(z)$.


Our analysis began by determining the global quantities $C$, ${\bar I}$, and $\Lambda$ along the equilibrium sequence of a given EOS, and constructing the profiles $X(z)$ of the corresponding stellar configurations along the sequence. The process was repeated for all the chosen EOS models. For the compactness $C$, we can construct a three-dimensional representation of the correlation space from the data set $(C, z, X(z))$. 
We only consider data for stable stars with masses ranging from $1 M_\odot$ up to the maximum mass configuration along each sequence.
For each stellar configuration, the profile $X(z)$ is constructed by using 40 chosen sampling points $z_i$ from the numerical solution. 
Similarly, we can construct the correlation spaces $({\bar I}, z, X(z))$ and $(\Lambda, z, X(z))$ for the 
dimensionless moment of inertia and tidal deformability. 
The data for the three correlation spaces are plotted in the first row of Fig.~\ref{fig:3d_plots}. 
As can be seen from the results, the data points exhibit a pronounced clustering in the vicinity of a
well-defined smooth surface in each correlation space. The results show that the profile $X(z)$ is weakly dependent on the EOSs for a given NS quantity $C$, $\bar I$ or $\Lambda$.


To quantitatively characterize the correlations, we implement a parametric representation through high-order polynomial fitting. We introduce the transformed coordinates $u \equiv 1 - z$ (with $u=0$ corresponding to the stellar surface) and $\xi$ stands for the NS quantity $C$, $\ln {\bar I}$ or $\ln \Lambda$. The data $X$ in each of the correlation spaces in Fig.~\ref{fig:3d_plots} are fit to a
two-dimensional surface $X(u, \xi)$ modeled by an eighth-order bivariate polynomial:
\begin{equation}
X(u, \xi) = \sum_{k=1}^{8} \sum_{m=0}^{8-k} c_{km} u^k \xi^m , 
\label{eq:fit}
\end{equation}
where the fitting coefficients $c_{km}$ are to be determined. This functional form explicitly satisfies the physical surface boundary condition $X(u=0, \xi)=0$, which is consistent with the vanishing of the 
pressure $P$ at the surface. 
The eighth-order expansion provides sufficient degrees of freedom to capture nonlinear features in the parameter space while maintaining computational tractability.

The optimal coefficients $c_{km}$, determined through least-squares regression over our ensemble of
star models constructed by the 45 chosen EOSs, are presented in Table~\ref{tab:coefficients} in 
Appendix~\ref{sec:EOS_models}.
To assess the accuracy of the fitting surfaces, we evaluate the root-mean-square error $\sigma$ for
each correlation defined by 
\begin{equation}
    \sigma = \sqrt{ {1\over N} \sum_{i=1}^N (X_i - {\hat X}_i) } , 
\end{equation}
where $N$ is the total number of data used in the fitting, $X_i$ is the exact value of $X$ in the data, and ${\hat X}_i$ is the approximated value of $X$ inferred from the corresponding fitting surface. 
We obtain $\sigma = 7.62 \times 10^{-3}$, $6.02\times 10^{-3}$, and $6.38\times 10^{-3}$, respectively, for the $(C, z, X(z))$, $({\bar I}, z, X(z))$, and $(\Lambda, z, X(z))$ correlations. 
With this analytical representation, the radial profile of $X(u, \xi=\xi_0)$, and hence that of the trace anomaly $\Delta$, for a NS with a measured quantity $\xi_0$ can be determined approximately.

Let us now focus on the value of the trace anomaly at the center of the star $\Delta_c$. 
In the second row of Fig.~\ref{fig:3d_plots}, we plot the data of $\Delta_c$ against the corresponding NS quantity in each panel. Specifically, the left panel shows the correlation between $\Delta_c$ and the compactness $C$, which is obtained from a slice of data at $z\equiv r/R=0$ in the 3D data plot $(C, z, X(z))$ as shown in the third row of Fig.~\ref{fig:3d_plots}. 
The blue lines in these two panels result from a fit to the $X_c$-$C$ data obtained by setting $u=1$ (for the star center) in Eq.~(\ref{eq:fit}). 
To evaluate the level of EOS sensitivity, we plot the relative difference $\Delta X_c/X_c$ between the numerical data of $X_c$ and the fitting curve in the fourth row of Fig.~\ref{fig:3d_plots}.
It can be seen that $X_c$ depends weakly on the EOS models and the relative difference is within 10\% for a
large range of $C$. 
This strong correlation between $X_c$ and $C$ has also been studied recently in \cite{Cai_2025}. 
In our study, we find that the correlation can be extended to the profile of $X$.   

Our data show that $\Delta_c$ passes through zero in the range of $C \sim 0.26-0.28$ and can reach 
$\Delta_c \approx -0.2$ near the maximum mass limits for some EOS models\footnote{It is not appropriate to define the relative difference between $\Delta_c$ and its corresponding fitting curve since $\Delta_c$ can cross zero. We thus use $\Delta X_c/X_c$ as a measure of the EOS sensitivity at the center.}.
Similarly, we plot $\Delta_c$ ($X_c$) against $\ln{\bar I}$ and $\ln{\Lambda}$ in the middle and right panels,
respectively, in the second (third) row of Fig.~\ref{fig:3d_plots}. 
The relations between $\Delta_c$ and these two quantities also depend weakly on the EOS models. The relative differences $\Delta X_c/X_c$ between the fits and the data are slightly smaller than that of the $X_c-C$ relation. The largest relative differences are about 9\% in these cases.

For a given NS quantity such as $\Lambda$, we can predict the profile of $X$ to within about 10\%. The uncertainty is due to the remaining EOS variation of Eq.~(\ref{eq:fit}). 
As we shall discuss in Sec.~\ref{sec:apply}, these NS quantities $C$, $\bar I$, and $\bar \Lambda$ can be
obtained from NS observations. The correlations we have established can be applied to estimate the profile of
trace anomaly from observations. 




\subsection{Validation tests}

To illustrate the accuracy of the analytical representation established in Eq.~(\ref{eq:fit}), we 
performed validation tests using five different EOS models, three of which are also chosen 
from the CompOSE database\footnote{We follow the EOS names used on the CompOSE website (https://compose.obspm.fr/)}: 
KBH(QHC21\_BT)~\cite{TOGASHI201778,Kojo_2022}, 
APR(APR) unified crust~\cite{akmalEquationStateNucleon1998,davisInferenceNeutronstarProperties2024,haenselEnvelopesStrongMagnetic2007}, and XMLSLZ(PKL1)~\cite{PhysRevC.69.034319,Xia_2022,PhysRevC.69.034319,niu2025propertiesmicroscopicstructuresdense}. 
Note that KBH(QHC21\_BT) is one of the 45 models used in the original fitting dataset. 
Besides these tabular EOS models, we also consider phenomenological models incorporating 
non-monotonic behavior of the speed of sound that differ significantly from typical hadronic EOSs~\cite{Tews_2018,McLerran_2019,Altiparmak_2022,Tan_2022,Annala_2023}.
Following the framework proposed in~\cite{Tews_2018}, we extend a low-density EOS by introducing
the following skewed Gaussian parametrization for the speed of sound squared $c_s^2$ as a function of the baryon number density $n$ at high densities (see Eq.~(9) of \cite{Tews_2018}): 
\begin{widetext}
    \begin{equation}
        c_s^2 = \frac{1}{3} - c_1 \exp\left[ - \frac{(n - c_2)^2}{n_{\rm BL}^2} \right]  
    + h_{\rm P} \exp\left[ - \frac{(n - n_{\rm P})^2}{w_{\rm P}^2} \right] \left( 1 + \text{erf}\left[ s_{\rm P} \frac{(n - n_{\rm P})}{w_{\rm P}} \right] \right),
    \label{eq:cs2_param}
    \end{equation}
\end{widetext}
where $n_{\rm BL}$ determines the transition baseline width. The peak of the speed of sound is characterized by its height $h_{\rm P}$, central position $n_{\rm P}$, width $w_{\rm P}$, and a skewness shape parameter $s_{\rm P}$. The coefficients $c_1$ and $c_2$ are uniquely determined by enforcing the continuity of both $c_s^2$ and its first derivative at the transition density where this high-density parametrization model connects to the low-density crust EOS, which is 
chosen to be the RG(SKI4) model \cite{REINHARD1995467,KOHLER1976301,PhysRevC.92.055803}. 
The transition density between the parameterized model and the low-density EOS is set at the nuclear saturation density. 
We select two representative parameter sets, designated as Para A and Para B, to test 
the accuracy of Eq.~(\ref{eq:fit}). The parameter values for Para A (Para B) are 
$n_{\rm BL}=1.5$ (2.0), $h_{\rm P}=0.43$ (0.40), $n_{\rm P} = 4.0$ (5.0), 
$w_{\rm P} = 2.0$ (1.5), and $s_{\rm P}=2.0$ (2.0), where 
$n_{\rm BL}$, $n_{\rm P}$ and $w_{\rm P}$ are given in units of the nuclear saturation number density $n_0$. 

\begin{figure}[H]
    \centering
    \begin{subfigure}[b]{0.5\textwidth}
        \centering
        \includegraphics[width=\textwidth]{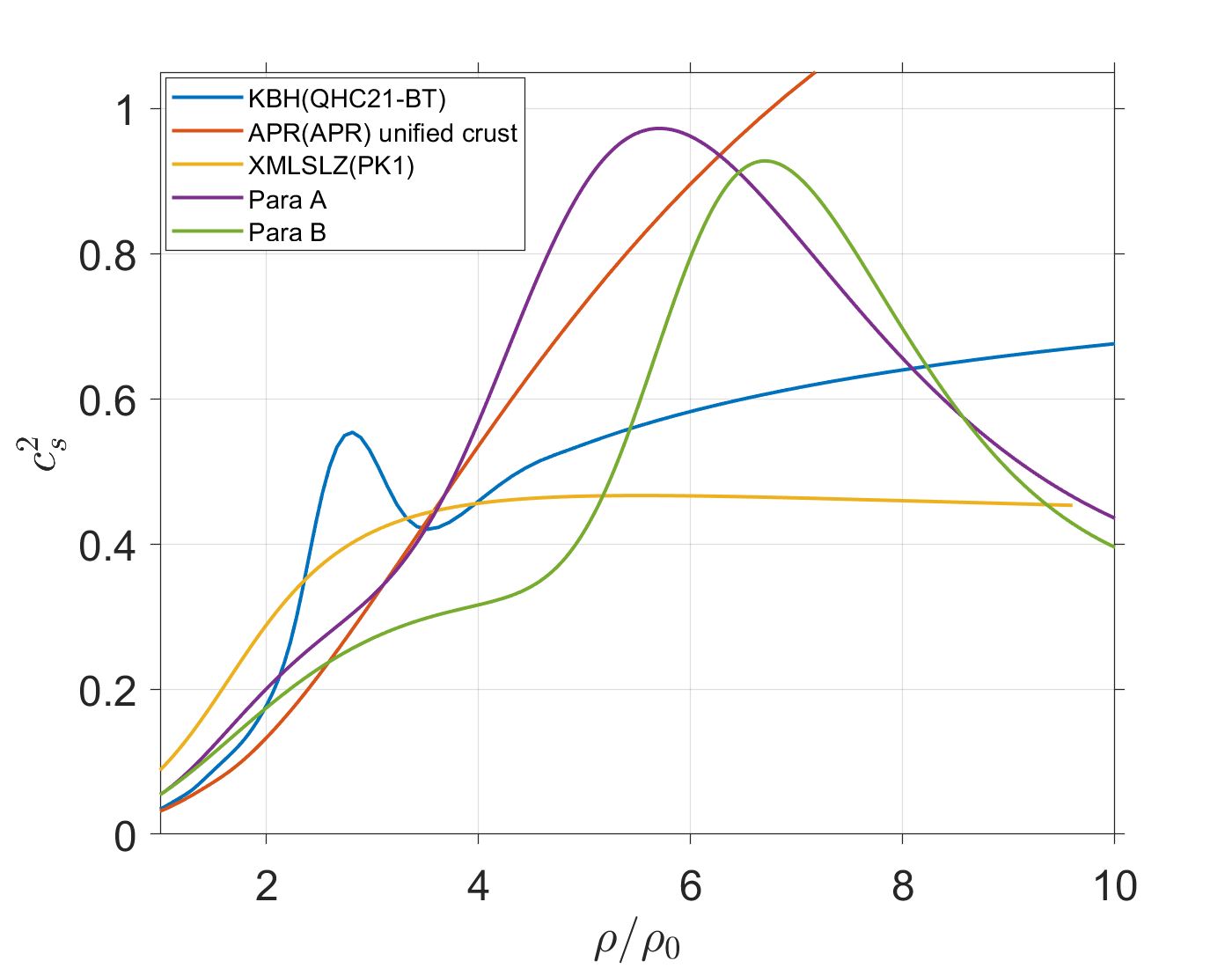}
    \end{subfigure}
    \vspace{1em}  
    \begin{subfigure}[b]{0.5\textwidth}
        \centering
        \includegraphics[width=\textwidth]{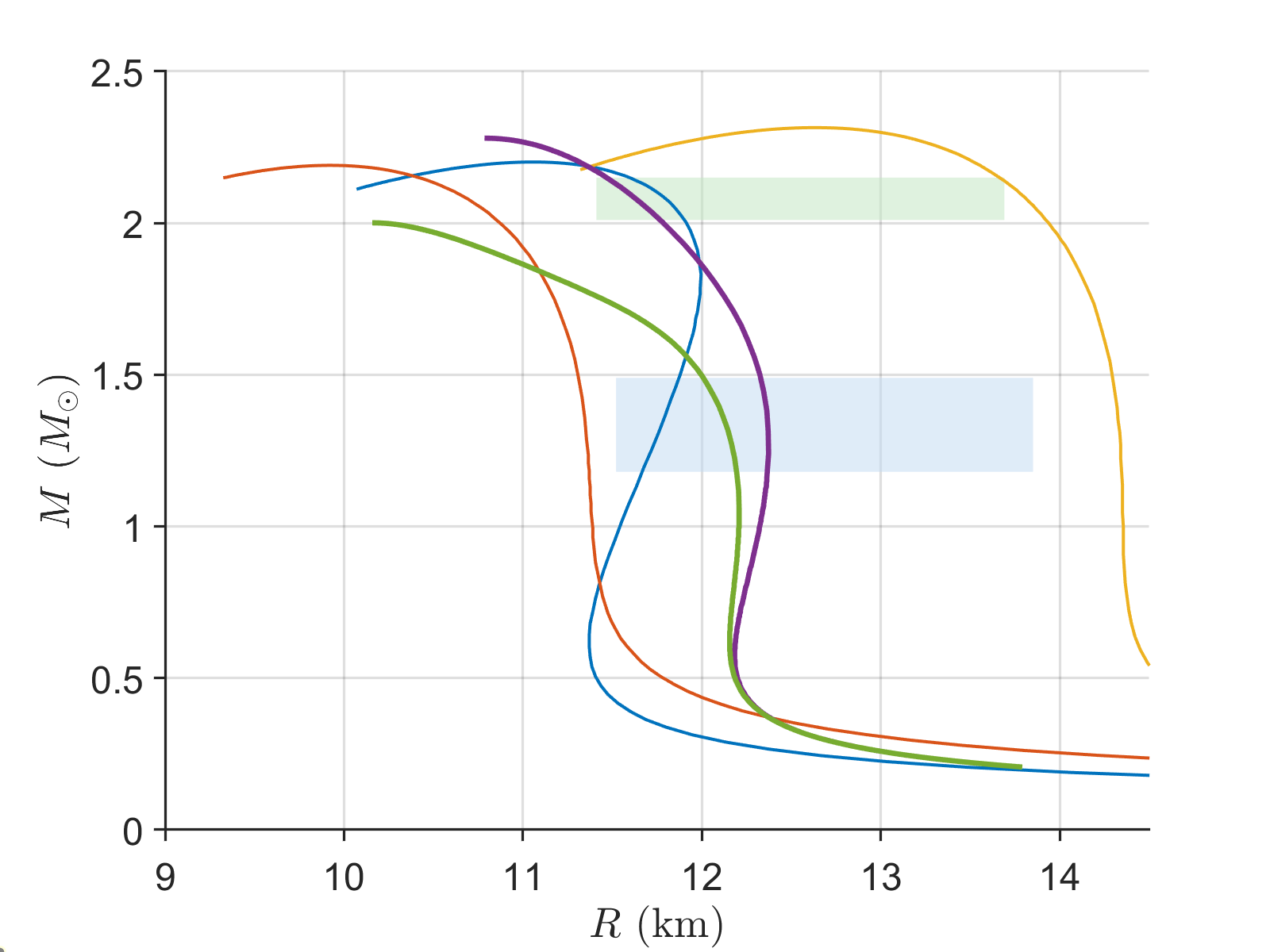}
    \end{subfigure}
    \caption{Upper panel: Speed of sound squared $c_s^2$ as a function of energy density $\rho$ (normalized by the nuclear saturation density $\rho_0$) for five EOS models used to test 
    Eq.~(\ref{eq:fit}). Lower panel: Mass-radius relations of these EOS models (shown by the same color lines).
    In contrast to the EOSs presented in Fig.~\ref{fig:MRtest}, some of these test models are not required to be compatible with the observational estimates for PSR J0740+6620 (green box) and PSR J0030+0451 (blue box).        } 
    \label{fig:test_models}
\end{figure}

In the upper panel of Fig.~\ref{fig:test_models}, we plot $c_s^2$ as a function of the normalized energy density $\rho/\rho_0$ for the five test models, where $\rho_0$ is the nuclear saturation energy density.  
The behaviors of $c_s^2$ for these chosen models are quite different. Specifically, the parameterized models Para A and Para B exhibit significant non-monotonic variations that deviate 
substantially from the three tabular EOS models.   
The peak of $c_s^2$ at intermediate density in the parameterized models is used to model the possibility of a crossover transition from nuclear to quark 
matter~\cite{Tews_2018, McLerran_2019}. 
The mass-radius relations of these models are also plotted in the lower panel of 
Fig.~\ref{fig:test_models} for comparison. Same as Fig.~\ref{fig:MRtest}, the rectangular boxes represent the observational estimates for PSR J0740+6620~\cite{Riley_2021} and 
PSR J0030+0451~\cite{Riley_2019}. 
In contrast to the 45 EOS models shown in Fig.~\ref{fig:MRtest} that are used to fit  
Eq.~(\ref{eq:fit}), the test models are deliberately chosen so that some of them only satisfy 
one observation data, but not both. One of the models is even incompatible with both data.

\begin{figure}[H]
    \includegraphics[width=\columnwidth]{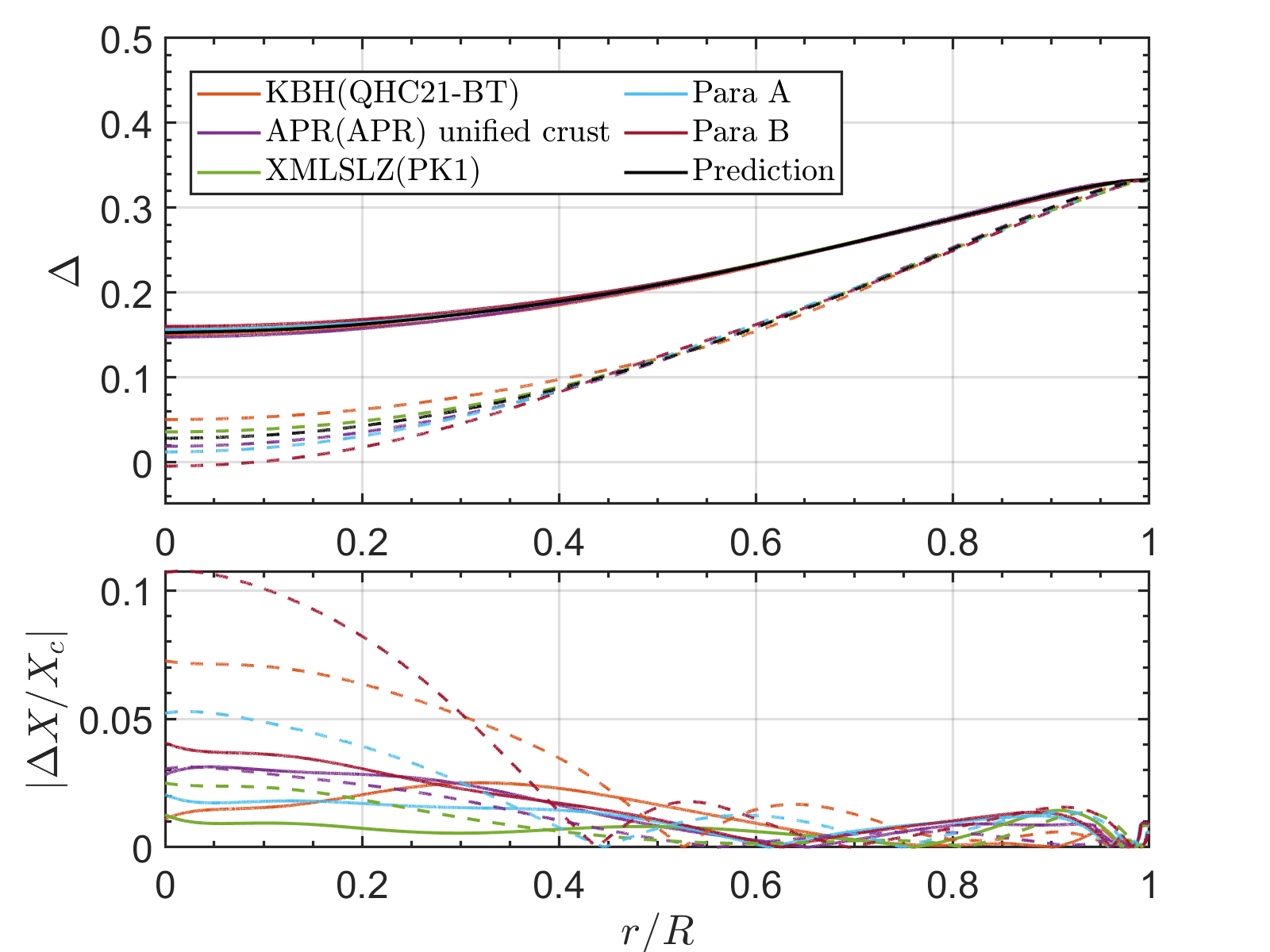}
    \caption{Upper panel: Profiles of $\Delta$ for five EOS models, four of which were not used in fitting 
    Eq.~(\ref{eq:fit}). Two values of compactness, $C=0.2$ (solid lines) and $0.26$ (dashed lines), are considered. The predictions of 
    Eq.~(\ref{eq:fit}) are represented by the black solid and dashed lines. 
    Lower panel: The errors $\Delta X/X_c$ between the EOS data and the predictions of Eq.~(\ref{eq:fit}).     }
\label{fig:XC_pred}
\end{figure}

We first test the performance of Eq.~(\ref{eq:fit}) in reconstructing the profile of the trace anomaly
$\Delta = 1/3 - X$ for a given compactness $C$, and assess the level of EOS sensitivity in the result. 
In the upper panel of Fig.~\ref{fig:XC_pred}, we plot the profiles of $\Delta$ for these test EOSs corresponding to compactness $C=0.2$ (solid lines) and $C=0.26$ (dashed lines).  
For the same compactness, it is seen that the profiles are very close to each other and can be modeled approximately by Eq.~(\ref{eq:fit}), which are represented by the black solid and dashed lines.

The variation due to different EOS models becomes noticeable in the inner region of the higher compactness stars.  
To quantify the errors in the prediction of Eq.~(\ref{eq:fit}), and thus the level of EOS sensitivity,
we normalize the absolute difference $\Delta X$ between the numerical data of $X$ and Eq.~(\ref{eq:fit}) by the predicted central value of $X_c$. The errors are shown in the lower panel of Fig.~\ref{fig:XC_pred}. At the center of the star $r=0$, $\Delta X/X_c$ is the same as the relative error and is within about 5\% for most of the data. 
The largest error at about 10\% is due to the high compactness $C=0.26$ star constructed by 
the parametrized model Para B. As shown in Fig.~\ref{fig:test_models}, this model does not satisfy 
the PSR J0740+6620 data. 
For the other parametrized model Para A, which satisfies both PSR J0740+6620 and PSR J0030+0451 data, 
the error is only about 5\% although this model also has a non-monotonic behavior of $c_s^2$. 
It should be noted that we define the error by normalizing the absolute error $\Delta X$ by $X_c$ 
instead of the corresponding value of $X$ along the radius. This is due to the fact that $\Delta X/X$ is not well defined at the surface $r=R$ where $X=0$. 
However, as seen in the upper panel of Fig.~\ref{fig:XC_pred}, the different profiles of $\Delta$
agree very well in the outer region of the stars for the same compactness. The profiles become more sensitive to the EOS only in the inner region. 

\begin{figure}[H]
    \includegraphics[width=\columnwidth]{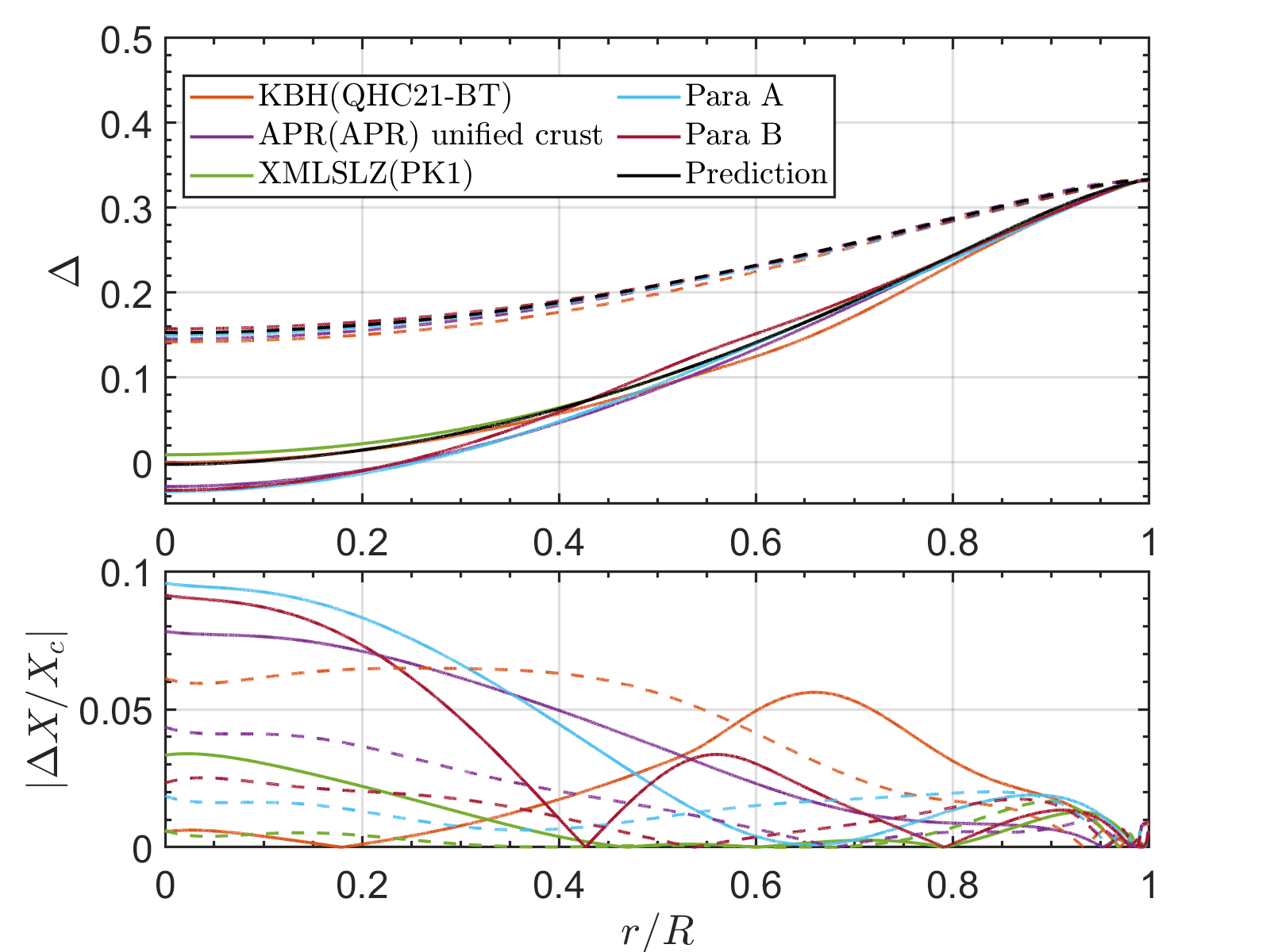}
    \caption{Similar to Fig.~\ref{fig:XC_pred}, but for two values of normalized tidal deformability
    $\ln \Lambda = 2.7$ (solid lines) and 5 (dashed lines). }
\label{fig:XL_pred}
\end{figure}

We have also used the five EOS models to check that Eq.~(\ref{eq:fit}) 
performs well for the other two NS quantities $\ln {\bar I}$ and $\ln \Lambda$. In Fig.~\ref{fig:XL_pred}, we show the profiles of $\Delta$ for $\ln \Lambda =2.7$ (solid lines) and $\ln\Lambda=5$ (dashed lines). 
The predictions of Eq.~(\ref{eq:fit}) are represented by the black lines. 
The profiles for different EOSs agree very well for $\ln \Lambda = 5$, corresponding to typical compactness values $C \approx 0.2$. The profiles become more sensitive to the EOSs for $\ln \Lambda=2.7$, corresponding to a higher compactness in the range about $C \approx 0.26-0.28$.
Nevertheless, Eq.~(\ref{eq:fit}) can still approximate the profiles of $\Delta$ quite well even for this case, with the error $\Delta X/X_c$ generally within 10\%. 
We also noted that $\Delta$ can drop below zero near the center of these highly compact NSs for some EOS models. 

The above validation tests demonstrate the applicability of Eq.~(\ref{eq:fit}) for EOS models not used 
in the original fitting.
We believe that the correlations between $\Delta$, $C$, $\ln {\bar I}$, and $\ln \Lambda$ presented in Fig.~\ref{fig:3d_plots} are generally true for NS EOS models that satisfy the observational constraints
presented in Fig.~\ref{fig:MRtest}. This conclusion should also apply to models incorporating 
non-monotonic behavior of $c_s^2$ that differ significantly from typical hadronic EOSs.
Eq.~(\ref{eq:fit}) can provide a simple and practical formula to determine the profile of $\Delta$ approximately when one of the three NS observables is given.

\section{Applications}
\label{sec:apply}

As an illustration to demonstrate the application of Eq.~(\ref{eq:fit}), we consider three different cases where the relevant NS quantities are either directly observed or inferred indirectly from specific NS systems. 

\subsection{Compactness}
\label{sec:apply_C}

We first apply Eq.~(\ref{eq:fit}) to the mass-radius measurements obtained by NICER, from which the compactness $C$ of the corresponding NSs can be determined. 
The observation for PSR J0030+0451 yields $M = 1.34_{-0.16}^{+0.15}~M_{\odot}$ and 
$R =12.71_{-1.19}^{+1.14}~\mathrm{km}$ \cite{Riley_2019}, resulting in the corresponding compactness within the range $C = 0.156^{+0.035}_{-0.030}$. 
Similarly, the measured values $M = 2.072^{+0.067}_{-0.066}~M_{\odot}$ and $R = 12.39^{+1.30}_{-0.98}~\mathrm{km}$ for the more massive star PSR J0740+6620 \cite{Riley_2021} give a significantly higher compactness $C = 0.247^{+0.030}_{-0.031}$.

\begin{figure}[H]
    \centering
    \includegraphics[width=\linewidth]{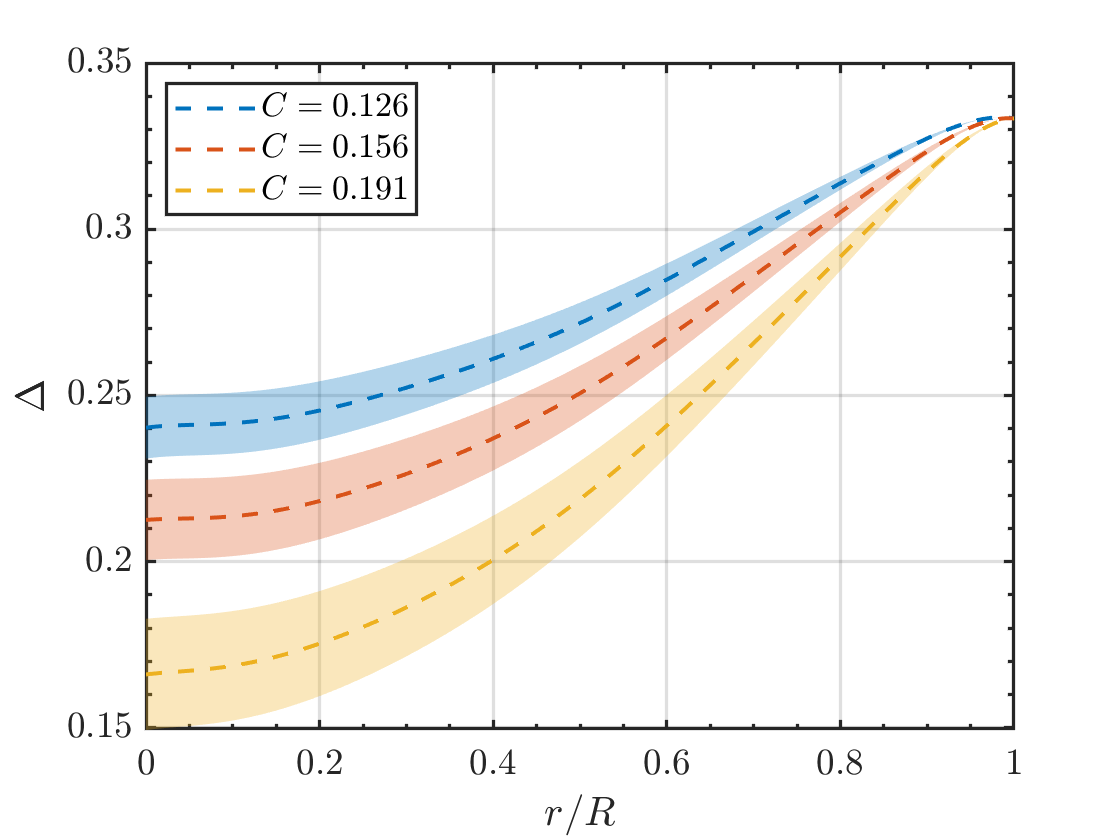}    
    \caption{The trace anomaly profile predicted by Eq.~(\ref{eq:fit}) according to the allowed range of 
    compactness for PSR J0030+0451 inferred from the mass-radius measurements. The red dashed line represents the profile for the best estimate value $C=0.156$, while the profiles inferred from the upper and lower bounds are represented by the other two dashed lines. The colored band around each 
    line represents an estimated level of $\pm 10\%$ uncertainty of Eq.~(\ref{eq:fit}) due to EOS sensitivity.}
    \label{fig:nicer2}
\end{figure}

\begin{figure}[H]
    \centering
    \includegraphics[width=\linewidth]{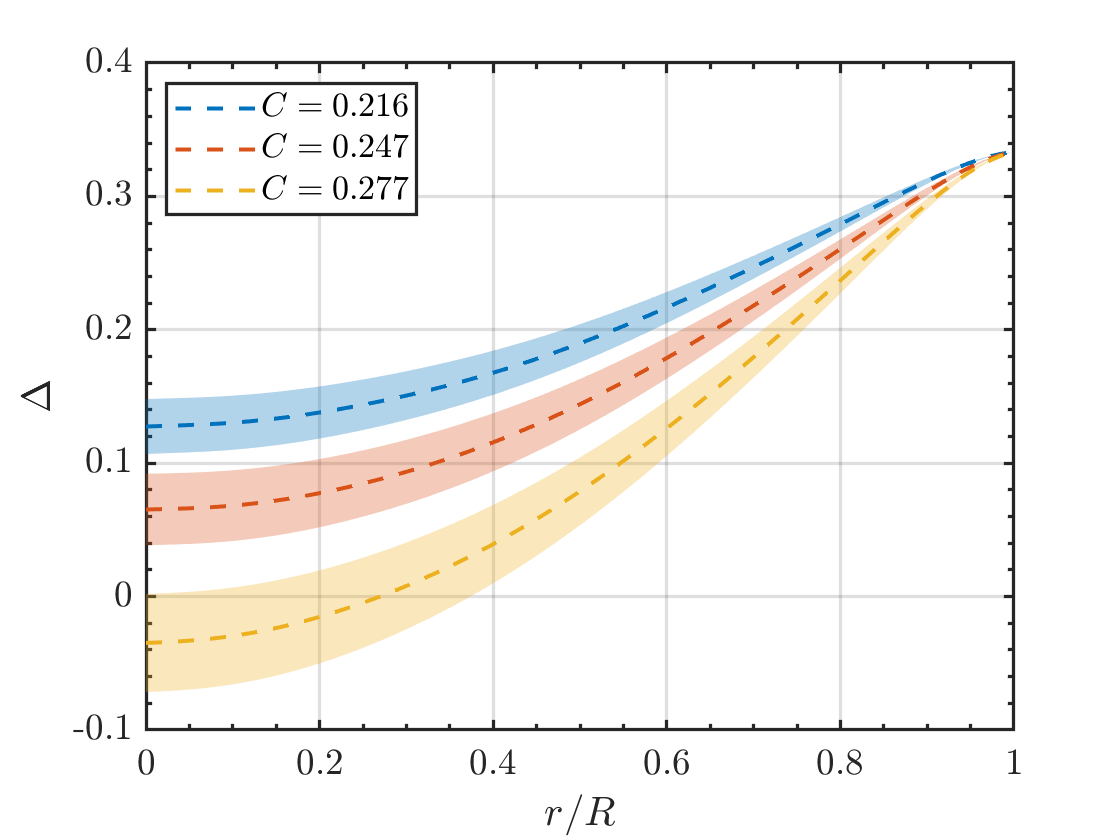}
    \caption{Similar to Fig.~\ref{fig:nicer2}, but for the more massive system PSR J0740+6620.}
    \label{fig:nicer1}
\end{figure}

For a given compactness $C$, we calculate the trace anomaly profile by $\Delta(u) = {1/3} - X(u, C)$, where $X(u,C)$ is determined by Eq.~(\ref{eq:fit}). 
In Fig.~\ref{fig:nicer2}, we show the profiles of $\Delta$ according to the observationally inferred 
interval $C = 0.156^{+0.035}_{-0.030}$ for the compactness of PSR J0030+0451. 
Specifically, the dashed lines represent the results for the three values $C=0.126$, 0.156, and 0.191. 
The colored band around each line represents an estimated level of $\pm 10\%$ uncertainty due to the EOS sensitivity in the function $X$. 
We also plot the corresponding results for the more massive star PSR J0740+6620 in Fig.~\ref{fig:nicer1}.  
Specifically, the central trace anomaly for PSR J0030+0451 and PSR J0740+6620 is estimated to be 
$\Delta_c = 0.2126^{+0.0370}_{-0.0630}$ and $\Delta_c = 0.0653^{+0.0828}_{-0.1368}$, respectively. 
These central values are consistent with those obtained recently in \cite{Cai_2025}. 
We also note that the possibility of having $\Delta_c < 0$, which contradicts the conjecture proposed in \cite{fujimotoTraceAnomalySignature2022}, for PSR J0740+6620 cannot be ruled out by the current observation data. 
Future X-ray missions, such as the enhanced X-ray Timing and Polarimetry mission (eXTP) \cite{AngLi_eXTP2025}, 
are expected to provide better mass-radius constraints and hence information on the trace anomaly. 

\subsection{Moment of inertia}

As mentioned in Sec.~\ref{sec:introduction}, the moment of inertia of PSR J0737-3039A is expected to be measurable to about 10\% accuracy in the near future \cite{lattimerConstrainingEquationState2005,Hu_Huanchen2020}. 
So far, there are various estimates for the moment of inertia of this NS 
(e.g., \cite{PhysRevC.100.035802,Kramer_2021, Silva_2021, Miao_2022}). 
In \cite{PhysRevC.100.035802}, deriving from Bayesian posterior probability distributions of the nuclear EOS
that incorporate information from many-body theory and empirical data of finite nuclei, the moment of 
inertia of PSR J0737-3039A with mass \(1.338 M_\odot\) is constrained to be in the range 
$1.04\times 10^{45} {\rm g\ cm}^2 < I < 1.51\times 10^{45} {\rm g\ cm}^2$ at the 95\% credibility level.
The most probable value is found to be $1.36\times 10^{45} {\rm g\ cm}^2$. This translates into the range for the dimensionless moment of inertia $10.046 < \bar{I} < 14.586$, with the most probable value 
$\tilde{\bar{I}} = 13.137$. For a given $\bar I$, the trace anomaly profile is determined by
$\Delta(u) = 1/3- X(u, \ln {\bar I})$.


\begin{figure}[H]
    \centering
    \includegraphics[width=\linewidth]{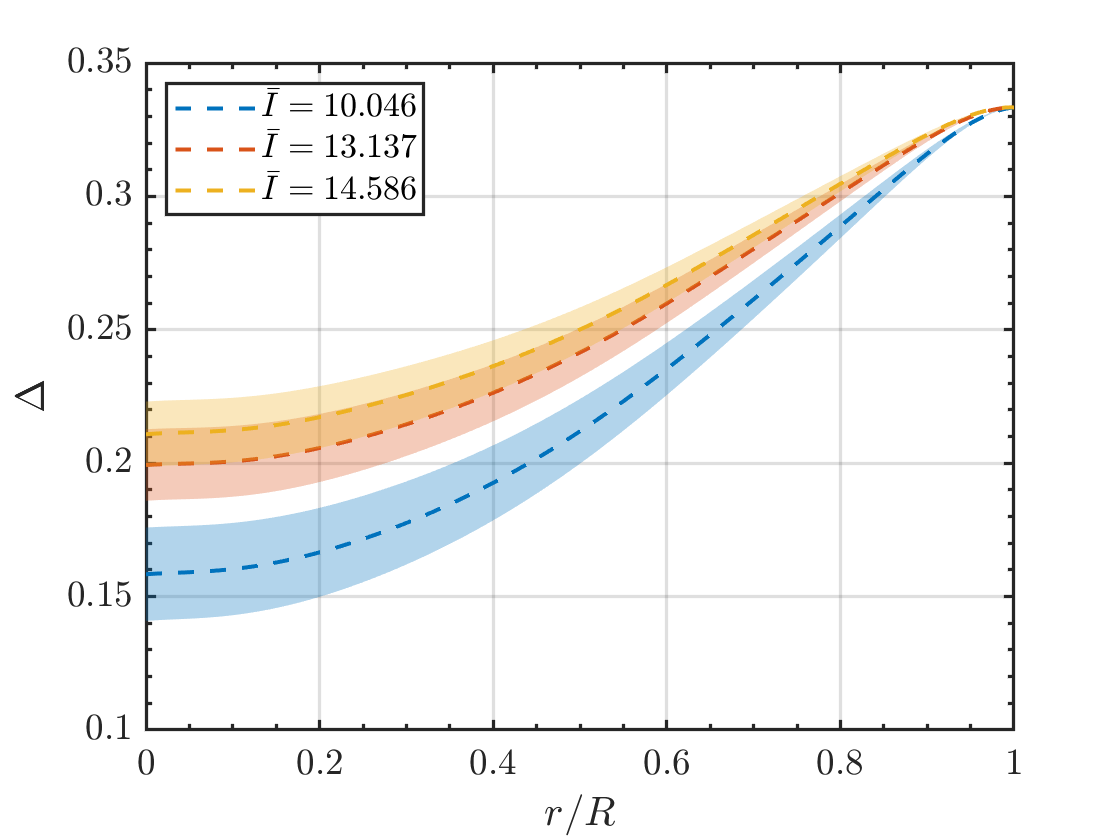}
    \caption{Similar to Fig.~\ref{fig:nicer2}, but for the range of dimensionless moment of inertia $\bar I$
    of PSR J0737-3039A obtained from the analysis of \cite{PhysRevC.100.035802}.}
    \label{fig:I_profile}
\end{figure}

Based on the inferred range of $\bar I$, we plot the corresponding trace anomaly profiles for PSR J0737-3039A
predicted by Eq.~(\ref{eq:fit}) in Fig.~\ref{fig:I_profile}. The red dashed line is the profile corresponding to the most probable value $\tilde{\bar{I}} = 13.137$, while the upper and lower bounds are represented by the  
other two dashed lines. Similar to Fig.~\ref{fig:nicer2}, the colored bands stand for an estimated $\pm 10\%$
uncertainty of Eq.~(\ref{eq:fit}). In particular, we obtain an estimate for the central trace anomaly to be $\Delta_c = 0.1993^{+0.0238}_{-0.0584}$. 
It is interesting to note that the mass of PSR J0030+0451 studied in Sec.~\ref{sec:apply_C} coincides 
with that of PSR J0737-3039A to about 10\%. The best estimated values of the central trace anomaly for these two systems, $\Delta_c = 0.2126$ and 0.1993, also agree to about 7\%.  


\begin{figure}[H]
    \centering
    \includegraphics[width=\linewidth]{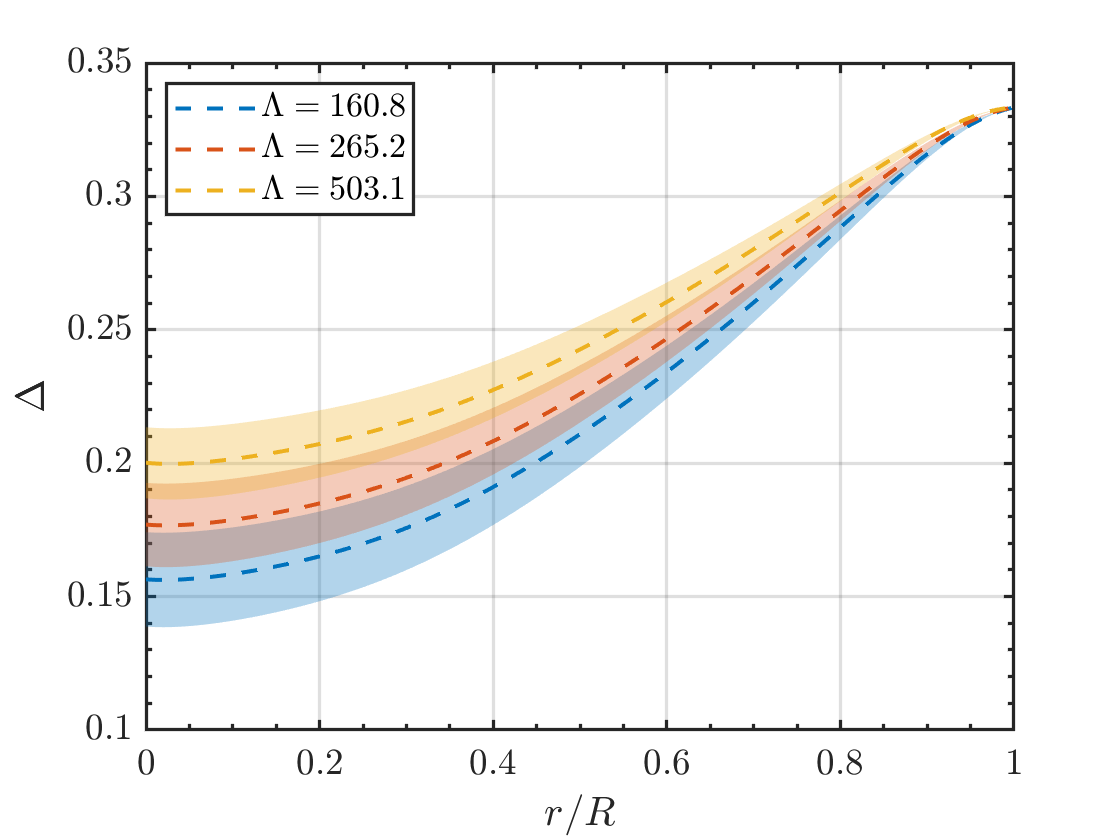}
    \caption{Similar to Fig.~\ref{fig:nicer2}, but for the range of dimensionless tidal deformability of
    a $1.4 M_\odot$ NS obtained from the analysis of \cite{Huang:2025mrd}.}
    \label{fig:placeholder}
\end{figure}

\subsection{Tidal deformability}

While the above investigations focus on NS quantities measurable from electromagnetic wave observations, 
the first gravitational wave signal from a binary NS system GW170817 \cite{abbottGW170817ObservationGravitational2017} has opened up a new channel 
to probe the properties of NSs. Specifically, various constraints on the tidal deformability of NSs have been obtained (e.g., \cite{GW170817_radius2018,LimHolt_2018,Essick_2020,Legred_2021}). In a recent work, Huang \cite{Huang:2025mrd} obtained an EOS-independent constraint on the dimensionless tidal deformability of a $1.4 M_\odot$ NS, $ \Lambda_{1.4}= 265.18^{+275.88}_{-104.38}$, by combining gravitational-wave-based inferences with NICER observations of PSR J0030+0451~\cite{Miller_nicer2019,Riley_2019} and 
PSR J0437-4715~\cite{Choudhury_2024}.        
Based on the inferred range of $\Lambda$, the trace anomaly profile is determined by 
$\Delta(u)=1/3 - X(u, \ln\Lambda)$ using Eq.~(\ref{eq:fit}). 
We plot the inferred profiles for a $1.4 M_\odot$ NS in Fig.~\ref{fig:placeholder}. The dashed lines are the results inferred from the best estimated value (red line), the upper and lower bounds. Similar to Fig.~\ref{fig:nicer2}, the colored bands represent an estimated $\pm 10\%$ uncertainty of Eq.~(\ref{eq:fit}). 
The central trace anomaly of a $1.4 M_\odot$ is determined to be $\Delta_c = 0.1770^{+0.0365}_{-0.0432}$.
We note that this estimated interval overlaps with the results obtained in \cite{Ecker_2023}, which studies the
dependence of the trace anomaly on the maximum mass of a NS.

There are other constraints derived for $\Lambda_{1.4}$ from previous studies as mentioned above. We could 
potentially repeat the analysis for comparison. However, this does not make a fundamental difference in this paper. Our aim is to use an example to demonstrate how Eq.~(\ref{eq:fit}) can be applied to observation data at this stage of our study. Currently, the error bars on $\Lambda_{1.4}$ from various studies remain relatively large. More precise and informative constraints on the trace anomaly are anticipated with the detection of more gravitational-wave events from binary NS mergers in the future by Advanced LIGO, Advanced Virgo, KAGRA, and 
third-generation detectors such as the Einstein Telescope and Cosmic Explorer.

\begin{figure}[H]
    \centering
    \includegraphics[width=\linewidth]{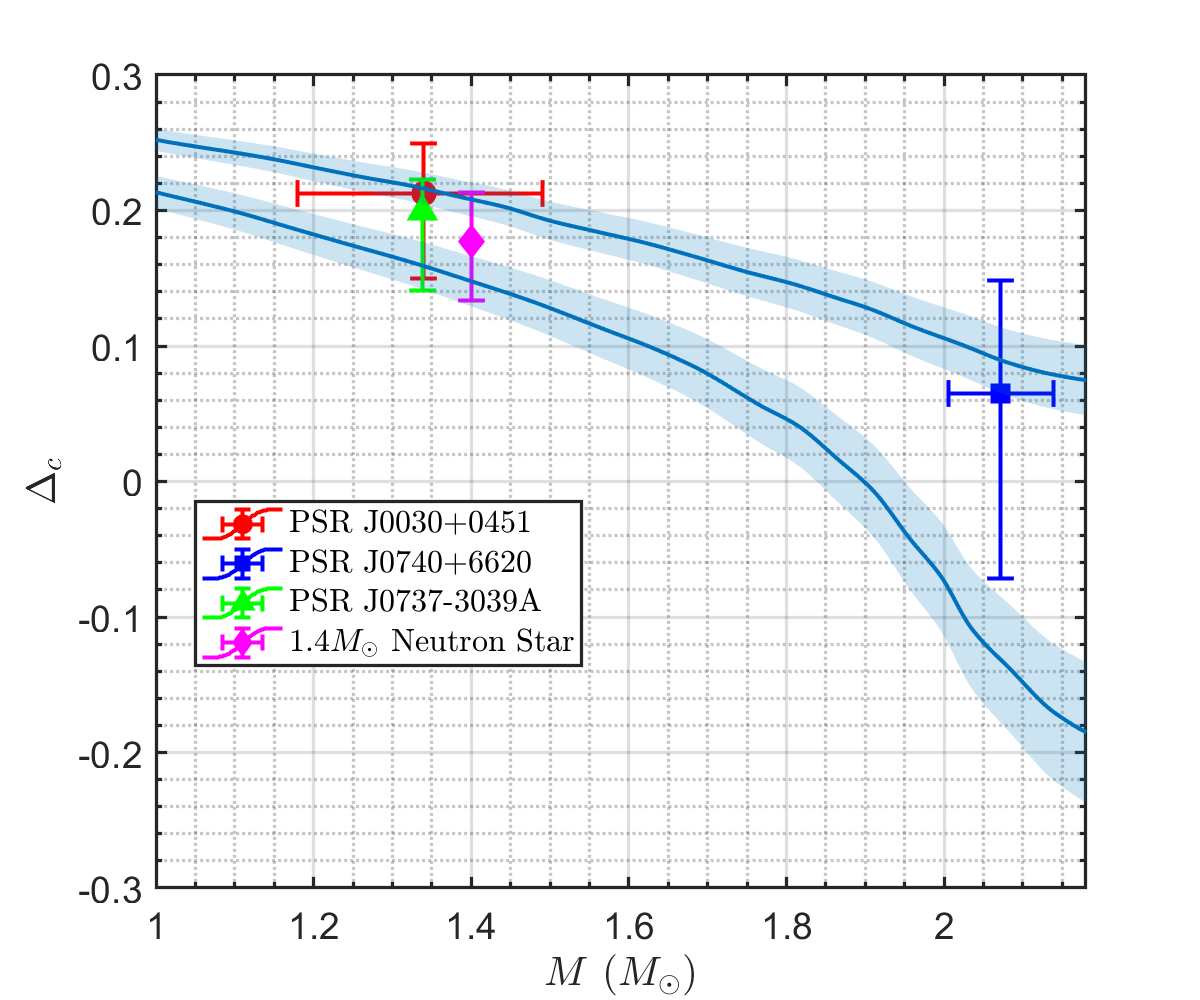}
    \caption{The central trace anomaly $\Delta_c$ as a function of stellar mass $M$ for the three pulsars PSR J0030+0451, PSR J0740+6620, PSR J0737-3039A, and a $1.4M_\odot$ canonical NS inferred in our analysis. 
    The blue sold lines are inferred from the boundaries of the interval on the tidal deformability-mass
    plane constrained by the observations of heavy pulsars and gravitational-wave signals in 
    \cite{Chatziioannou2020} (see text).   }
    \label{fig:Delta_c_all}
\end{figure}

\section{Discussion}
\label{sec:discuss}

In this work, we have explored the correlations between the trace anomaly $\Delta$ and three important observables of NSs: compactness $C$, dimensionless moment of inertia ${\bar I}\equiv I/M^3$, and 
dimensionless tidal deformability $\Lambda \equiv \lambda/M^5$. By selecting 45 EOS models based on various nuclear theories that 
satisfy the mass-radius measurements of PSR J0740+6620 and PSR J0030+0451, we found that the profile 
of $X\equiv P/\rho$, which serves as a proxy for the trace anomaly $\Delta = 1/3-X$ in this study, is weakly 
dependent on the EOS model for a given C, $\bar I$ or $\Lambda$. We generate an ensemble of NS configurations and fit the data with an eighth-order bivariate polynomial (see Eq.~(\ref{eq:fit})), resulting in three 
quasi-universal relations that relate the profile of $X$ separately to $C$, $\bar I$, and $\Lambda$, as shown in Fig.~\ref{fig:3d_plots}.

To illustrate the application of our results, we first used the compactness values inferred from the mass-radius measurements of PSR J0740+6620 and PSR J0030+0451 to determine the trace anomaly profiles for these NSs. 
We then used the estimate values for the moment of inertia of PSR J0737-3039A from the analysis of 
\cite{PhysRevC.100.035802} to infer the trace anomaly profiles of this NS. This system is particularly 
interesting as its moment of inertia is expected to be measurable to about 10\% accuracy in the near future \cite{lattimerConstrainingEquationState2005,Hu_Huanchen2020}. 
Furthermore, we applied the EOS-independent multimessenger constraint on the dimensionless tidal deformability,
as determined in \cite{Huang:2025mrd}, to derive the trace anomaly profile of a $1.4 M_\odot$ canonical NS 
model. 

For comparison, we plot the inferred central trace anomaly $\Delta_c$ for all the cases mentioned above against 
the stellar mass in Fig.~\ref{fig:Delta_c_all}. Although there is an estimate $10\%$ uncertainty of Eq.~(\ref{eq:fit}) due to EOS sensitivity, the error bars in $\Delta_c$ are currently dominated by observational errors. 
We found that the possibility of $\Delta_c < 0$, which contradicts the conjecture proposed in \cite{fujimotoTraceAnomalySignature2022}, at the center of PSR J0740+6620 cannot be ruled out by current
observation data. More precise mass-radius measurements by future X-ray missions will potentially 
resolve this matter. 

In Fig.~7 of \cite{Chatziioannou2020}, a 90\% credible interval on the tidal deformability-mass plane is shown
based on the constraints from heavy pulsars (with mass $\sim 2 M_\odot$) and two binary NS gravitational-wave signals, GW170817 and GW190425 (see the purple region in Fig.~7 of \cite{Chatziioannou2020}). 
We extracted the upper and lower bounds of that interval and translated them into the blue solid lines on 
the $\Delta_c-M$ plane shown in Fig.~\ref{fig:Delta_c_all} using Eq.~(\ref{eq:fit}). The blue band around
each line represents an estimated $\pm 10\%$ uncertainty of Eq.~(\ref{eq:fit}) due to EOS sensitivity. 

We see that the two solid lines enclose well the four data points derived separately from $C$, $\bar I$,
and $\Lambda$ in our analysis. 
It is interesting to note that the $1.4 M_\odot$ NS data point lies well between the two boundaries. This 
alignment may stem from the fact that this data point is inferred also from the gravitational-wave constraints on the tidal deformability, and thereby reassuring the consistency of our analysis. For the other three data points inferred from the compactness and moment of inertia, which are determined independently of 
gravitational-wave constraints, their best estimated values lie closer to the upper solid line. 

In summary, our main results are effectively shown by Fig.~\ref{fig:Delta_c_all}, showcasing  
multimessenger information on $\Delta_c$ derived from the latest NS observations available to date through 
both electromagnetic and gravitational-wave channels.  
Future data from these channels is expected to give tighter constraints, offering valuable insights into how dense matter inside neutron stars approaches the conformal limit or not.

We conclude this paper with a few remarks. (1) Although we have only chosen 45 EOS models to establish the quasi-universal relations (Eq.~(\ref{eq:fit})), we believe our results generally apply to EOSs which satisfy the current mass-radius measurements for PSR J0740+6620 and PSR J0030+0451 as shown in Fig.~\ref{fig:MRtest}. However, it should be noted that the maximum NS masses allowed by the chosen EOSs cover only a range of about 2 to $2.5 M_\odot$. If more massive NSs ($\gtrsim 2.5 M_\odot$) are discovered in the future, our quasi-universal relations may require a revision by incorporating additional EOS models capable of supporting such massive NSs. 
Nevertheless, future observations are expected to provide tighter constraints on mass-radius measurements and hence also help narrow the range of possible EOS models. 
(2) We have only considered traditional NS models; therefore, our quasi-universal relations 
may not be applicable to hybrid star models featuring a strong first-order phase transition. 
It has been shown in \cite{Jimenez_2024} that $\Delta$ suffers a sharp decrease around the phase transition point, potentially breaking the quasi-universal relations. 
In fact, since the energy density is discontinuous in a first-order phase transition, the assumption
of a continuous $X$ specified by Eq.~(\ref{eq:fit}) is not valid. 
(3) The nature of our quasi-universal relations is qualitatively different from other NS universal relations, such as the I-Love-Q relations~\cite{yagiILoveQRelationsNeutron2013}. 
Instead of connecting different NS global quantities, our relations provide direct links between the 
NS observables and microphysics, enabling an estimation of the trace anomaly despite our ignorance of the nuclear matter EOS.
Our results also imply that $\Delta$ can become negative, as inferred from some neutron star data (albeit with relatively large observation error bars). This may indicate non-trivial properties of dense matter in NSs, such as color superconductivity~\cite{Fukushima_2025}. More precise NS observations in the future will help resolve this issue.

\section*{Acknowledgements}
SR acknowledges the support of Poling Class Scholarship at Nankai University and
the Visiting Student Program at the Chinese University of Hong Kong, where part of this work was conducted. LML is supported by a grant from the Research Grants Council of Hong Kong SAR, China (Project No: 14304322).   

\appendix

\section{EOS models and fitting coefficients}
\label{sec:EOS_models}

Table~\ref{tab:eos_list_row_sorted} lists the 45 EOS models chosen from the CompOSE database 
\cite{typelCompOSECompStarOnline2013} for fitting Eq.~(\ref{eq:fit}). We follow the EOS names used on the
CompOSE website (https://compose.obspm.fr/). Table~\ref{tab:coefficients} 
presents the numerical values of the fitting coefficients $c_{km}$ in Eq.~(\ref{eq:fit}). 

\begin{table*}[h!]
\centering
\caption{Alphabetized List of the nuclear EOS models used in fitting Eq.~(\ref{eq:fit}). }
\label{tab:eos_list_row_sorted}

\begin{tabular}{l l}
\toprule
CMGO (GDFM-II)~\cite{Carreau2019,PhysRevD.108.103045,PhysRevC.77.025802} & DNS(CMF) hadronic (cold neutron stars) with crust~\cite{Dexheimer_2008,Schürhoff_2010,PhysRevC.92.012801,Dexheimer_2017} \\ \addlinespace
DS(CMF)-1~\cite{Dexheimer_2008,Dexheimer_2017,PhysRevC.103.025808,PhysRevC.81.045201} & DS(CMF)-1 with crust~\cite{PhysRevC.40.2834,PhysRevC.92.055803,Dexheimer_2008,Dexheimer_2017,PhysRevC.81.045201,PhysRevC.103.025808} \\ \addlinespace
DS(CMF)-2~\cite{Dexheimer_2008,Dexheimer_2017,PhysRevC.81.045201,PhysRevC.103.025808} & DS(CMF)-2 with crust~\cite{PhysRevC.40.2834,PhysRevC.92.055803,Dexheimer_2008,Dexheimer_2017,PhysRevC.81.045201,PhysRevC.103.025808} \\ \addlinespace
DS(CMF)-4~\cite{Dexheimer_2008,Dexheimer_2017,PhysRevC.81.045201,PhysRevC.103.025808} & DS(CMF)-5 with crust~\cite{PhysRevC.40.2834,PhysRevC.92.055803,Dexheimer_2008,Dexheimer_2017,PhysRevC.81.045201,PhysRevC.103.025808} \\ \addlinespace
DS(CMF)-6~\cite{Dexheimer_2008,Dexheimer_2017,PhysRevC.81.045201,PhysRevC.103.025808} & DS(CMF)-7~\cite{Dexheimer_2008,Dexheimer_2017,PhysRevC.81.045201,PhysRevC.103.025808} \\ \addlinespace
DS(CMF)-7 with crust~\cite{PhysRevC.40.2834,PhysRevC.92.055803,Dexheimer_2008,Dexheimer_2017,PhysRevC.81.045201,PhysRevC.103.025808} & DS(CMF)-8~\cite{Dexheimer_2008,Dexheimer_2017,PhysRevC.81.045201,PhysRevC.103.025808} \\ \addlinespace
DS(CMF)-8 with crust~\cite{PhysRevC.40.2834,PhysRevC.92.055803,Dexheimer_2008,Dexheimer_2017,PhysRevC.81.045201,PhysRevC.103.025808} & GDTB(DDHdelta)~\cite{GAITANOS200424,PhysRevC.90.045803,refId0} \\ \addlinespace
GPPVA(DD2) NS unified inner crust-core~\cite{PhysRevC.81.015803,PhysRevC.90.045803,10.1093/mnras/stz800} & GPPVA(FSU2) NS unified inner crust-core~\cite{PhysRevC.90.045803,providenciaHyperonicStarsNuclear2019,PhysRevC.90.044305,10.1093/mnras/stz800} \\ \addlinespace
GPPVA(FSU2R) NS unified inner crust-core~\cite{PhysRevC.90.045803,PhysRevC.90.044305,10.1093/mnras/stz800,Negreiros_2018,providenciaHyperonicStarsNuclear2019} & KBH(QHC21\_A)~\cite{TOGASHI201778,Kojo_2022} \\ \addlinespace
KBH(QHC21\_AT)~\cite{TOGASHI201778,Kojo_2022} & KBH(QHC21\_B)~\cite{TOGASHI201778,Kojo_2022} \\ \addlinespace
KBH(QHC21\_BT)~\cite{TOGASHI201778,Kojo_2022} & KBH(QHC21\_C)~\cite{TOGASHI201778,Kojo_2022} \\ \addlinespace
KBH(QHC21\_DT)~\cite{TOGASHI201778,Kojo_2022} & OPGR(DDHdeltaY4) (with hyperons)~\cite{GAITANOS200424,Oertel_2015,PhysRevC.90.045803,refId0} \\ \addlinespace
OPGR(GM1Y5) (with hyperons)~\cite{PhysRevLett.67.2414,Oertel_2015,refId0} & OPGR(GM1Y6) (with hyperons)~\cite{PhysRevLett.67.2414,Oertel_2015,refId0} \\ \addlinespace
PCGS(PCSB0)~\cite{HEMPEL2010210,PhysRevC.98.065804,PRADHAN2023122578} & PCGS(PCSB1)~\cite{HEMPEL2010210,PhysRevC.98.065804,PRADHAN2023122578} \\ \addlinespace
PCP(BSK22)~\cite{Audi_2017,universe7120470,PhysRevC.105.015803,PhysRevC.101.015802,10.1093/mnras/stz800,PhysRevC.88.061302,PhysRevC.100.035801,refId01,PhysRevLett.119.192502} & PCP(BSK24)~\cite{Audi_2017,universe7120470,PhysRevC.105.015803,PhysRevC.101.015802,10.1093/mnras/stz800,PhysRevC.88.061302,PhysRevC.100.035801,refId01,PhysRevLett.119.192502} \\ \addlinespace
PT(GRDF2-DD2) cold NS~\cite{Typel2014,Typel_2018,doi:10.1142/9789813209350_0004} & RG(Rs)~\cite{DANIELEWICZ200936,PhysRevC.33.335,PhysRevC.92.055803} \\ \addlinespace
RG(SK255)~\cite{DANIELEWICZ200936,PhysRevC.68.031304,PhysRevC.92.055803} & RG(SK272)~\cite{DANIELEWICZ200936,PhysRevC.68.031304,PhysRevC.92.055803} \\ \addlinespace
RG(SKa)~\cite{DANIELEWICZ200936,KOHLER1976301,PhysRevC.92.055803} & RG(Skb)~\cite{DANIELEWICZ200936,KOHLER1976301,PhysRevC.92.055803} \\ \addlinespace
RG(SkI3)~\cite{REINHARD1995467,KOHLER1976301,PhysRevC.92.055803} & RG(SkI4)~\cite{REINHARD1995467,KOHLER1976301,PhysRevC.92.055803} \\ \addlinespace
RG(SkI6)~\cite{REINHARD1995467,KOHLER1976301,PhysRevC.92.055803} & SPG(M4) unified NS EoS~\cite{10.1093/mnras/stz800,PhysRevC.77.025802,PhysRevD.109.103015} \\ \addlinespace
SPG(M5) unified NS EoS~\cite{10.1093/mnras/stz800,PhysRevC.77.025802,PhysRevD.109.103015} & XMLSLZ(DD-LZ1)~\cite{Xia_2022,PhysRevC.105.045803,Wei_2020,niu2025propertiesmicroscopicstructuresdense} \\ \addlinespace
XMLSLZ(DDME-X)~\cite{Xia_2022,PhysRevC.105.045803,TANINAH2020135065,niu2025propertiesmicroscopicstructuresdense} & XMLSLZ(DDME2)~\cite{PhysRevC.71.024312,Xia_2022,PhysRevC.105.045803,niu2025propertiesmicroscopicstructuresdense} \\ \addlinespace
XMLSLZ(PKDD)~\cite{PhysRevC.71.024312,Xia_2022,PhysRevC.69.034319,niu2025propertiesmicroscopicstructuresdense} & \\ \addlinespace
\bottomrule
\end{tabular}
\end{table*}

\begin{table*}[htbp] 
\centering
\renewcommand{\arraystretch}{1.25} 
\caption{8th-order polynomial coefficients $c_{km}$ for Eq.~(\ref{eq:fit})}
\label{tab:coefficients}

\sisetup{
    round-mode=places,
    round-precision=3,
    scientific-notation=true
}

\begin{tabular}{c@{\hspace{10pt}}c@{\hspace{10pt}}c}


\begin{tabular}{c}
\multicolumn{1}{c}{\bfseries (a) $X-\mathcal{C}$ relation} \\
\begin{tabular}{|c|c|@{\hspace{10pt}}S[table-format=1.3e5]|}
\hline\hline
$k$ & $m$ & {$c_{km}$} \\
\hline
1 & 0 &  2.42565616864368 \\
  & 1 & -115.504146761970 \\
  & 2 &  2123.85557695614 \\
  & 3 & -20432.5009862913 \\
  & 4 &  112238.615344613 \\
  & 5 & -353158.678019447 \\
  & 6 &  592015.352940175 \\
  & 7 & -409730.874124871 \\
\hline
2 & 0 &  0.0866897318664307 \\
  & 1 &  50.9549443186949 \\
  & 2 & -865.238244354615 \\
  & 3 &  8112.13952783051 \\
  & 4 & -39524.5425696822 \\
  & 5 &  95982.3592274315 \\
  & 6 & -89133.7179814663 \\
\hline
3 & 0 & -5.99689623728736 \\
  & 1 & -4.86417290328816 \\
  & 2 & -324.186667078163 \\
  & 3 &  2018.07313250296 \\
  & 4 & -5714.62808210648 \\
  & 5 &  3630.91824157337 \\
\hline
4 & 0 &  20.1339772890549 \\
  & 1 &  75.6599237005761 \\
  & 2 & -95.8169654392024 \\
  & 3 &  308.560218249588 \\
  & 4 &  856.458379586535 \\
\hline
5 & 0 & -45.2671694981549 \\
  & 1 & -80.5566446272432 \\
  & 2 & -4.26443237329450 \\
  & 3 & -360.702200659608 \\
\hline
6 & 0 &  55.3328131757355 \\
  & 1 &  54.7561303756579 \\
  & 2 &  63.6764088996897 \\
\hline
7 & 0 & -35.2962112345838 \\
  & 1 & -20.6072681915500 \\
\hline
8 & 0 &  9.33366919987045 \\
\hline\hline
\end{tabular}
\end{tabular}

&

\begin{tabular}{c}
\multicolumn{1}{c}{\bfseries (b) $X-\ln {\bar I}$ relation} \\
\begin{tabular}{|c|c|@{\hspace{10pt}}S[table-format=1.3e5]|}
\hline\hline
$k$ & $m$ & {$c_{km}$} \\
\hline
1 & 0 &  125.488775358021 \\
  & 1 & -334.533222155577 \\
  & 2 &  377.905690438285 \\
  & 3 & -233.758292040519 \\
  & 4 &  85.3717607182069 \\
  & 5 & -18.3941452855634 \\
  & 6 &  2.16392343944564 \\
  & 7 & -0.107193118581190 \\
\hline
2 & 0 &  89.1286894460065 \\
  & 1 & -142.979292175691 \\
  & 2 &  94.3172069372224 \\
  & 3 & -30.2487743263134 \\
  & 4 &  4.30708090467563 \\
  & 5 & -0.0868137959241926 \\
  & 6 & -0.0253842177339250 \\
\hline
3 & 0 & -129.882120758942 \\
  & 1 &  165.594947888453 \\
  & 2 & -99.0085372487877 \\
  & 3 &  31.5876678104632 \\
  & 4 & -5.24294871296069 \\
  & 5 &  0.360458573597570 \\
\hline
4 & 0 &  120.563849964604 \\
  & 1 & -71.8827498937904 \\
  & 2 &  21.4221720529405 \\
  & 3 & -2.95979346128054 \\
  & 4 &  0.116794748204572 \\
\hline
5 & 0 & -128.232594926868 \\
  & 1 &  40.5117881138346 \\
  & 2 & -7.12965597983566 \\
  & 3 &  0.627234644422956 \\
\hline
6 & 0 &  98.1026983410868 \\
  & 1 & -14.0018769450283 \\
  & 2 &  0.632226657154620 \\
\hline
7 & 0 & -45.9758396500157 \\
  & 1 &  2.81384190002598 \\
\hline
8 & 0 &  9.33366919222049 \\
\hline\hline
\end{tabular}
\end{tabular}

&

\begin{tabular}{c}
\multicolumn{1}{c}{\bfseries (c) $X-\ln \Lambda$ relation} \\
\begin{tabular}{|c|c|@{\hspace{10pt}}S[table-format=1.3e5]|}
\hline\hline
$k$ & $m$ & {$c_{km}$} \\
\hline
1 & 0 &  0.279286943240886 \\
  & 1 & -0.00753891785870781 \\
  & 2 & -0.0312134913299423 \\
  & 3 &  0.0164051701108091 \\
  & 4 & -0.00495032354548114 \\
  & 5 &  0.000791789037217683 \\
  & 6 & -6.21361808313499e-05 \\
  & 7 &  1.89176238678700e-06 \\
\hline
2 & 0 &  7.30308688815337 \\
  & 1 & -3.22283834988232 \\
  & 2 &  0.722558684459142 \\
  & 3 & -0.0777216995544163 \\
  & 4 &  0.00267308533769101 \\
  & 5 &  0.000164615897534155 \\
  & 6 & -1.20923496233335e-05 \\
\hline
3 & 0 & -18.2589072081838 \\
  & 1 &  6.21405742970747 \\
  & 2 & -1.15442665276940 \\
  & 3 &  0.116243898139230 \\
  & 4 & -0.00660556282213797 \\
  & 5 &  0.000191561540841853 \\
\hline
4 & 0 &  22.8507728878466 \\
  & 1 & -5.74235266775833 \\
  & 2 &  0.749426112091682 \\
  & 3 & -0.0389625266734431 \\
  & 4 &  0.000268326439860235 \\
\hline
5 & 0 & -14.0183051505511 \\
  & 1 &  2.89047374059891 \\
  & 2 & -0.302951001176694 \\
  & 3 &  0.0126576111576940 \\
\hline
6 & 0 &  0 \\
  & 1 & -0.458223263844711 \\
  & 2 &  0.0210243355540088 \\
\hline
7 & 0 &  4.55444276966077 \\
  & 1 & -0.00371083599796490 \\
\hline
8 & 0 & -1.75891421105833 \\
\hline\hline
\end{tabular}
\end{tabular}

\end{tabular}


\end{table*}
	
\clearpage

\bibliography{bib}

\end{document}